\documentclass[twocolumn, tighten, dvipsnames, floatfix]{aastex63}

\usepackage[T1]{fontenc}
\usepackage[]{xcolor}
\usepackage{booktabs}
\usepackage{physics}
\usepackage{wasysym}
\usepackage{longtable}
\usepackage{threeparttablex} % for "ThreePartTable" 
\usepackage{tabularx}
\usepackage{graphicx}

\usepackage{array}
\usepackage{amsmath}
\usepackage{changepage}
\usepackage{lineno, ulem} 

\usepackage[flushmargin]{footmisc}
\usepackage{afterpage}
\usepackage{subfigure}
\usepackage{multirow}

\DeclareUnicodeCharacter{FF0C}{,}

\newcommand{\kms}{\mbox{$\>{\rm km\, s^{-1}}$}}

\def\arcsec{\hbox{$^{\hbox{\rlap{\hbox{\lower4pt\hbox{$\,\prime\prime$}}
          }}}$} \ }

\def\arcmin{\hbox{$^{\hbox{\rlap{\hbox{\lower4pt\hbox{$\;\prime$}}
          }\hbox{$\frown$}}}$}}

\usepackage{lipsum}
\usepackage{float}
\usepackage[misc]{ifsym}

\shorttitle{Early Co-formation of the Milky Way's Disks}
\shortauthors{Borbolato et al.}

\begin{document}

\title{Early Co-formation of the Milky Way's Thin and Thick Disks at Redshift $z > 2$}

\correspondingauthor{Lais Borbolato (laisborbolato@usp.br), Jo\~ao A. S. Amarante (joaoant@gmail.com) }
% \correspondingauthor{Jo\~ao A. S. Amarante}
% \email{email\_joão}

%Correto
\author[0000-0003-3382-1051]{Lais Borbolato}
\affil{Universidade de S\~ao Paulo, Instituto de Astronomia, Geof\'isica e Ci\^encias Atmosf\'ericas, Departamento de Astronomia, \\ SP 05508-090, S\~ao Paulo, Brasil}

%Correto
\author[0000-0001-7479-5756]{Silvia Rossi}
\affil{Universidade de S\~ao Paulo, Instituto de Astronomia, Geof\'isica e Ci\^encias Atmosf\'ericas, Departamento de Astronomia, \\ SP 05508-090, S\~ao Paulo, Brasil}

%Correto
\author[0000-0002-0537-4146]{H\'elio D. Perottoni}
\affil{Universidade de S\~ao Paulo, Instituto de Astronomia, Geof\'isica e Ci\^encias Atmosf\'ericas, Departamento de Astronomia, \\ SP 05508-090, S\~ao Paulo, Brasil}
\affil{Laborat\'{o}rio Nacional de Astrof\'{i}sica, Rua Estados Unidos 154, 37504-364, Itajub\'{a} - MG, Brazil}
\affil{Observat\'{o}rio Nacional, MCTI, Rua Gal. Jos\'{e} Cristino 77, Rio de Janeiro, 20921-400, RJ, Brazil}

%Correto
\author[0000-0002-9269-8287]{Guilherme~Limberg}
\affiliation{Kavli Institute for Cosmological Physics, University of Chicago, 5640 S Ellis Avenue, Chicago, IL 60637, USA}

%Correto
\author[0000-0002-7662-5475]{Jo\~ao A. S. Amarante}
\affil{Department of Astronomy, School of Physics and Astronomy, \\ Shanghai Jiao Tong University, 800 Dongchuan Road, Shanghai, 200240, China}
\affil{State Key Laboratory of Dark Matter Physics, School of Physics and Astronomy，\\ Shanghai Jiao Tong University, Shanghai, 200240, China}

%Correto
\author[0000-0001-9209-7599]{Anna B. A. Queiroz}
\affiliation{Instituto de Astrof\'isica de Canarias, E-38200 La Laguna, Tenerife, Spain}
\affiliation{Departamento de Astrof\'isica, Universidad de La Laguna, E-38205 La Laguna, Tenerife, Spain}

%Correto
\author[0000-0003-1269-7282]{Cristina~Chiappini}
\affiliation{Leibniz-Institut f\"ur Astrophysik Potsdam (AIP), An der Sternwarte 16, D-14482 Potsdam, Germany}

%Correto
\author[0000-0003-4524-9363]{Friedrich Anders}
\affiliation{Departament de Física Quàntica i Astrofísica (FQA), Universitat de Barcelona (UB), C Martí i Franquès, 1, 08028 Barcelona, Spain}
\affiliation{Institut de Ci\`encies del Cosmos, Universitat de Barceloana (ICCUB), C Mart\'i i Franqu\`es 1, 08028 Barcelona, Spain}
\affiliation{Institut d’Estudis Espacials de Catalunya (IEEC), Edifici RDIT, Campus UPC, 08860 Castelldefels (Barcelona), Spain}

\author[0000-0002-7529-1442]{Rafael M. Santucci}
\affil{Universidade Federal de Goi\'as, Instituto de Estudos Socioambientais, Planet\'ario, Goi\^ania, GO 74055-140, Brazil}
\affil{Universidade Federal de Goi\'as, Campus Samambaia, Instituto de F\'isica, Goi\^ania, GO 74001-970, Brazil}

\author[0000-0002-8262-2246]{Fabrícia O. Barbosa}
\affil{Universidade de S\~ao Paulo, Instituto de Astronomia, Geof\'isica e Ci\^encias Atmosf\'ericas, Departamento de Astronomia, \\ SP 05508-090, S\~ao Paulo, Brasil}

\author[0009-0007-5867-0583]{João V. Nogueira-Santos}
\affil{Universidade de S\~ao Paulo, Instituto de Astronomia, Geof\'isica e Ci\^encias Atmosf\'ericas, Departamento de Astronomia, \\ SP 05508-090, S\~ao Paulo, Brasil}

%%%%%%%%%%%%%%%% ABSTRACT %%%%%%%%%%%%%%%%%%%%

\begin{abstract}

The Milky Way serves as a template for understanding the formation and evolution of late-type massive disk galaxies since we can obtain detailed chemical and kinematic information for large samples of individual stars. However, the early formation of the disk and the dichotomy between the chemical thick and thin disks remain under intense debate. Some mechanisms have been proposed to explain the formation of this dichotomy, such as the injection of metal-poor gas by a gas-rich merger such as Gaia-Sausage Enceladus (GSE), or by cosmic gas filaments, radial migration, and the presence of star-forming clumps at high redshift ($z > 2$). In this work, we combine astrometric data from the \textit{Gaia} mission, chemical abundances from APOGEE and LAMOST spectroscopic surveys, and \texttt{StarHorse} ages to map the evolution of our Galaxy. The Bayesian isochrone-fitting code \texttt{StarHorse} can estimate ages for thousands of stars in the solar neighborhood, being most reliable for main sequence turnoff and sub-giants, computing distances and extinction simultaneously. From these samples, we show that (i) there is an old thin disk population ($>$11 Gyr) that indicates a period of co-formation between the thick and thin disks of the Milky Way before the GSE merger, i.e. the Galaxy itself could initiate the formation of a low-$\alpha$ disk without the need for a gas-rich merger, and (ii) this merger would have been important to stop the formation of stars in the thick disk.

\end{abstract}


%%%%%%%%%%%%%%%%%%%%%%%%%%%%%%%%%%%%%%%%%%%%%%%%%%
%%%%%%%%%%%%%%%% INTRODUCTION %%%%%%%%%%%%%%%%%%%%
%%%%%%%%%%%%%%%%%%%%%%%%%%%%%%%%%%%%%%%%%%%%%%%%%%

\section{Introduction}
\label{sec:intro}

Within the broad context of galaxy evolution, the Milky Way (MW) provides the best available laboratory for studying the detailed properties, on a star-by-star basis, of massive spirals, something not yet possible for distant galaxies. In the current era, we are able to obtain detailed chemical and kinematic information for millions of individual stars in our Galaxy in a way that was never before possible. This is thanks to recent photometric (e.g., \citealt{york2000, skrutskie2006, Chambers2016, Starkenburg2017}), and spectroscopic (e.g., \citealt{steinmetz2002, SEGUE1, LAMOST2012, LAMOST2012b, de2015galah, APOGEE2017, SEGUE2}) surveys, but especially astrometric data from the \textit{Gaia} mission \citep{GaiaMission}, which provides reliable parallax and proper motion measurements for almost two billion stars. Since it is well known that the chemical composition observed in the atmospheres of stars reflects the composition of the environment of their birth (e.g., \citealt{Freeman2002, BlandHawthorn2016}), we can use this combination of data to map the characteristics of the stellar populations present in the MW and reconstruct their formation and evolutionary history. 

Historically, the vertical density profile of the MW disk is well-described by a two exponential model with different scale heights (\citealt{Yoshii1982, Gilmore1983, Yoshii1987}). Each exponential represents the geometrical thick and thin disks where the former has a higher scale height \citep[$0.9\,{\rm kpc}$;][]{LiZhao2017} compared to the latter \citep[$0.3\,{\rm kpc}$;][]{Juric2008} , and occupies a region that is spatially less extended in Galactocentric radius and wider in height compared to the thin disk (e.g., \citealt{Bovy2012, Hayden2015, Sharma2021}). The differences between these components also extend to the kinematic properties; for example, the thick disk contains dynamically hotter stars (e.g., \citealt{Bond2010, Kordopatis2013}) with slightly more eccentric orbits \citep[$0.2 \lesssim e \lesssim 0.6$;][]{Villalobos2008, Sales2009} than the thin disk ($e < 0.2$). A major distinction in the MW disk is strongly delineated in the [$\alpha$/Fe]--[Fe/H] diagram, which exhibits two sequences (\citealt{Fuhrmann1998, Bensby2003, Lee2011, Fuhrmann2011, Anders2014, Hayden2015, Queiroz2020, Sharma2021, Imig2023}). The chemical thick disk appears mostly as a high-$\alpha$ sequence, more enriched in [$\alpha$/Fe] than the chemical thin disk at the same [Fe/H], which is associated with the low-$\alpha$ region. It is important to emphasize that the different definitions of thin and thick disks (e.g., geometrical, chemical, or kinematical) imply different properties. In this paper, we are interested mostly in the chemical differences.

An important ingredient in interpreting the [$\alpha$/Fe]--[Fe/H] diagram is the time delay between core-collapse supernovae (CCSNe) explosions and the Type Ia supernovae (SNeIa) in the Galaxy. The $\alpha$-elements are produced mainly by CCSNe, which correspond to the final stages of the evolution of massive stars and take a few Myr to occur \citep{Burrows2021} after star formation. In contrast, Fe is produced in greater quantities by SNeIa, which depends on the explosions of white dwarfs in binary systems, a process that can take up to 1\,Gyr \citep{Howell2011}. This delay in the Fe enrichment timescale causes a decrease (``\textit{knee}'' in [$\alpha$/Fe]--[Fe/H] diagram, \citealt{Tinsley1980, Matteucci1986}) in [$\alpha$/Fe] \citep{Matteucci2021}. The extent of the [Fe/H] of low-$\alpha$ stars at each position is a combination of stars born in different parts of the Galaxy, which can be triggered by internal secular processes (e.g., \citealt{Minchev2013, Anders2017}).

In addition to the well-known chemical and kinematic characteristics of the high- and low-$\alpha$ populations, it is firmly established that the high-$\alpha$ population is predominantly composed of very old stars (e.g., \citealt{Fuhrmann2011, Haywood2013, Anders2018}). The age of the high-$\alpha$ population was estimated to be around 11\,Gyr by \citet{Miglio2021}, with 95\% of this population forming within approximately 1.5\,Gyr, making them nearly coeval. These age estimates were derived from asteroseismic data on giant stars obtained from Kepler and spectra from the Apache Point Observatory Galactic Evolution Experiment (APOGEE; \citealt{APOGEE2017}) survey. This age was further confirmed by \citet{Queiroz2023}, who applied the isochrone-fitting method using the \texttt{StarHorse} code (\citealt{Santiago2016starhorse, Queiroz2018, Queiroz2023}) to main-sequence turnoff (MSTO) and sub-giant branch (SGB) stars, utilizing data from different spectroscopic surveys. In contrast, the mean age of the low-$\alpha$ population was found to be significantly younger (e.g., \citealt{Lagarde2021, Queiroz2023, Nepal2024a}).

With respect to the formation channel for the thin and thick-disk distinct stellar populations, \citet{Chiappini1997} propose two episodes of gas accretion to form the thick and thin disks, in this order or sequentially (\citealt{Chiappini2009IAU, Grisoni2017}), the two-infall model (see also \citealt{Spitoni2019, Spitoni2021, Spitoni2023}). The gas injected during a second infall would be responsible for diluting the elemental abundances in the interstellar medium and decreasing the star formation rate for a few Myr. This decreases the rates of CCSNe, but does not affect the SNIa explosions. As a result, the observed [$\alpha$/Fe] decreases, leading to the formation of the extended low-$\alpha$ sequence. In this context, there are different forms of gas accretion, e.g. cosmological simulations show that cold gas can be accreted through a filamentary structure (e.g. \citealt{AnglesAlcazar2017, Agertz2021, Renaud2021}). Another impactful form of gas accretion would be through a major wet merger. For the MW, the event responsible for the second infall could be, in principle, the merger of \textit{Gaia-Sausage Enceladus} (GSE; \citealt{belokurov2018, Haywood2018, helmi2018}). GSE is the dwarf galaxy responsible for the last major event of our Galaxy, dating to around 9 to 11\,Gyr (e.g., \citealt{Gallart2019, Bonaca2020, Montalban2021}).

The formation of the thin/low-$\alpha$ and thick/high-$\alpha$ disks is also a natural consequence of a clumpy phase of the evolution of a MW-like galaxy. In this scenario, intermediate-mass clumps form in the disk in the first $\sim$3 Gyr of evolution and they give rise to the chemical and geometrical disks (\citealt{Clarke2019, Beraldo2021}) with similar kinematics when compared to the MW disk \citep{Amarante2020}. The thick disk would have formed in regions of high gas density, known as clumps, with high levels of $\alpha$-elements enrichment due to their high star formation efficiency. Meanwhile, distributed continuous star formation produces the thin disk (low-$\alpha$ sequence). During this first Gyr of evolution, clumps will sink towards the Galactic center \citep{Garver2023} and also give rise to the bulge's chemical bimodality \citep{Debattista2023}. A critical prediction of this scenario is the co-formation of the thin and thick disks as shown in \citet{Beraldo2021}. They showed that an isolated clumpy galaxy has a non-negligible fraction of old low-$\alpha$ stars in agreement with observations of APOGEE and ages from \citet{Sanders&Das2018} for a sample of MSTO and giants stars. This co-formation scenario for thin and thick disks was supported by observations of RR Lyrae with high azimuthal motions and low-velocity dispersion, i.e., a dynamically cold configuration \citep[][]{Prudil2020, DOrazi2024}. Since RR Lyrae stars are ubiquitously very old ($>$ 10\,Gyr), this detection was interpreted as evidence of an early formation of the thin disk via clumps, with its oldest stars having ages comparable to the high-$\alpha$/thick disk population \citep{Beraldo2021}. The presence of very old metal-rich stars with kinematic characteristics coinciding with the thin disk population was also reported by \citet{Nepal2024} using spectra from the \textit{Gaia} Radial Velocity Spectrometer, \textit{Gaia}-RVS \citep{Guiglion2024} and spectro-photometric \texttt{StarHorse} ages. 

%%%%%%%%%%% FIGURA 1 %%%%%%%%%%%%
\begin{figure*}[ht!]
    \centering
    \includegraphics[width=2.1\columnwidth]{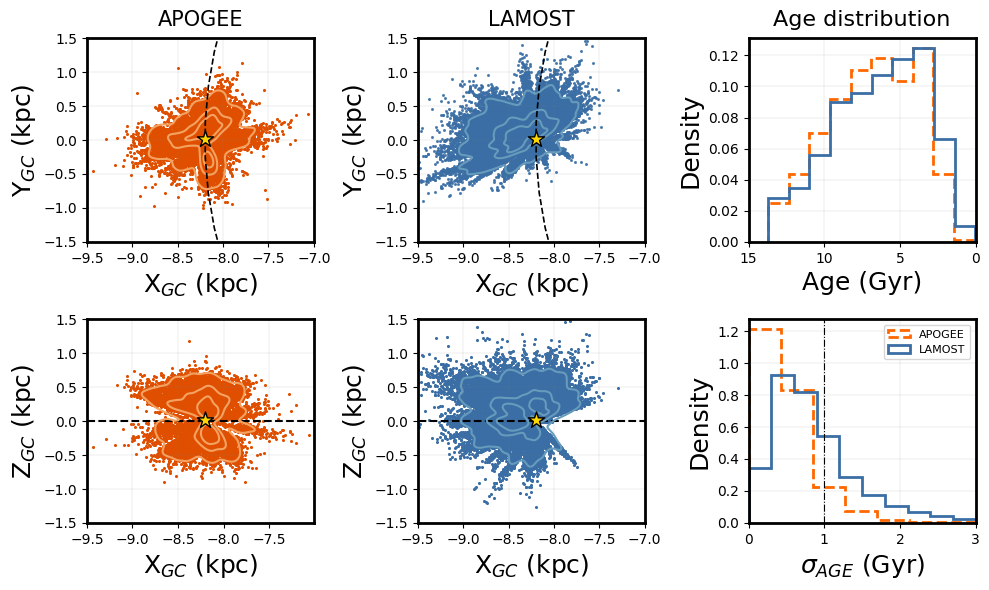}
    \caption{Spatial distribution of the $Y_{GC}$--$X_{GC}$ (top panels) and $Z_{GC}$--$X_{GC}$ (bottom panels) planes of the final APOGEE DR17 and LAMOST MRS DR7 samples in first and second columns, respectively. The yellow star indicates the position of the Sun ($[X, Y, Z]_{\rm GC} = [-8.2, 0.0, 0.0]$\,kpc). In the top panels, the dashed lines indicate the Galactocentric circumference for the radius of the Sun, and in the bottom panels, the dashed line indicates the Galactic plane ($Z_{\rm GC}$ = 0\,kpc). The top panel of the rightmost column shows the age distribution of the entire samples, considering the quality cuts. The bottom row of the same column shows the uncertainty in age ($\sigma_{AGE}$) of samples, the vertical black dashed line represents the cut in $\sigma_{AGE}$ = 1\,Gyr.}
    \label{spatial_dist}
\end{figure*}

Thanks to recent spectroscopic surveys and stellar parallaxes measured with \textit{Gaia}, the \texttt{StarHorse} code has provided ages and distances for millions of stars in the Galaxy with an isochrone-fitting method \citep{Queiroz2023}. Hence, this data set presents a great opportunity to take a closer look at stellar ages and revisit, from this point of view, the previously discussed scenarios regarding the MW disk formation. In this paper, we test the hypothesis of thin and thick disk co-formation at early times of the MW, prior to the GSE merger ($\gtrsim$11\,Gyr) with stellar ages for MSTO and SGB stars provided by the newest run of the \texttt{StarHorse} spectro-photometric code for the APOGEE DR17 and the Large Sky Area Multi-Object Fiber Spectroscopic Telescope (LAMOST; \citealt{LAMOST2012, LAMOST2012b}) DR7 medium-resolution (MRS) data. 

This paper is organized as follows. In Section \ref{sec:data}, we describe the data utilized in this work. Section \ref{sec:selection} details the method used to select candidate stars from the thin and thick disks. In Section \ref{sec:results} we present our results and discuss how they fit into the currently proposed scenarios in Section \ref{sec:discussion}. Finally, we summarize our results in Section \ref{conclusion}.

%%%%%%%%%%%%%%%%%%%%%%%%%%%%%%%%%%%%%%%%%%
%%%%%%%%%%%%%%%% DATA %%%%%%%%%%%%%%%%%%%%
%%%%%%%%%%%%%%%%%%%%%%%%%%%%%%%%%%%%%%%%%%

\section{Data}
\label{sec:data}

To obtain the detailed chemical composition of MW disk stars in the solar neighborhood, we selected stars from the APOGEE DR17 (\citealt{Abdurrouf2022}) and the LAMOST MRS DR7 spectroscopic surveys. The choice of two spectroscopic surveys aims to show that the observed trends go beyond the calibrations of the surveys themselves and avoid biases introduced by target-selection functions, increasing the reliability of our results. The chemical abundances and stellar parameters come from the spectroscopic surveys, and stellar distances and ages are derived by \citet{Queiroz2023} using \texttt{StarHorse}. Figure \ref{spatial_dist} shows the spatial distribution of the APOGEE and LAMOST samples, considering the quality cuts (see below), in $Y_{GC}$--$X_{GC}$ and $Z_{GC}$--$X_{GC}$ cartesian planes, and the age and age uncertainty ($\sigma_{AGE}$) distributions. 

\subsection{The \texttt{StarHorse} code}
\label{sec:SH}

The \texttt{StarHorse} code (\citealt{Santiago2016starhorse, Queiroz2018}) is a Bayesian isochrone fitting tool that is able to derive stellar parameters such as distances $d$, extinctions $A_v$ ($\lambda$ = 542 nm), ages $\tau$, masses $m_{\star}$, effective temperatures $T_{\text{eff}}$, metalicities [M/H], and surface gravities log $g$. These parameters are derived from measured stellar properties from spectroscopic and photometric surveys and Galactic priors for the main components of the MW. \citet{Anders2019, Anders2022} published distances determined using \textit{Gaia} parallaxes and photometry alone for a large dataset of $\sim$300 million stars. Afterward, a more comprehensive set of observables is also employed by incorporating spectroscopic atmospheric parameters in \citet{Queiroz2020, Queiroz2023}, allowing for the computation of stellar ages as well as more accurate and precise distances. The astrometric parameters adopted are from \textit{Gaia} DR3 \citep{GAIADR32023}. The set of observables is matched with stellar evolutionary models from PAdova and TRiestre Stellar Evolution Code (PARSEC; \citealt{Bressan2012}) to obtain the desired parameters.

\citet{Queiroz2023} derived the stellar parameters for eight samples of spectroscopic surveys (see Section 3 of their paper for more details), and for the first time with the \texttt{StarHorse} code, provided stellar ages for large datasets. The influence of the age prior adopted is discussed in Appendix \ref{sec:appendix}, in which we conclude that it does not alter the conclusions presented in this paper. The ages estimated are reliable for MSTO and SGB stars because isochrones of varying ages are best distinguished in this region of the Hertzprung-Rusell diagram. Unfortunately, the low luminosity of both MSTO and SGB stars leads to age samples being restricted to the solar neighborhood ($d < 2$\,kpc). For all the spectroscopic releases with parameters estimated by \citet{Queiroz2023}, APOGEE DR17 and the LAMOST MRS DR7 samples typically have the smallest nominal uncertainties. Having said that, we limited both APOGEE and LAMOST samples to stars with uncertainty in estimated age less than 1\,Gyr (vertical dashed black line in the lower rightmost panel of Figure \ref{spatial_dist}), and fractional distance uncertainty of $d/\sigma_{d} > 5$. We found that if the age uncertainty cut is removed, the qualitative results are maintained, including the fraction of thin and thick disk stars for ages older than 11\,Gyr. However, we choose to maintain the selection $\sigma_{\text{AGE}} \leq 1$\,Gyr to better highlight our results. Since \texttt{StarHorse} utilized astrometric data from \textit{Gaia}, we implemented the selection criteria based on the recommended range of renormalized unit weight errors ($\texttt{RUWE} \leq 1.4$; \citealt{Lindegren2020a}) to ensure high-quality astrometry. 

\begin{figure*}[ht!]
    \centering
    \includegraphics[width=2.1\columnwidth]{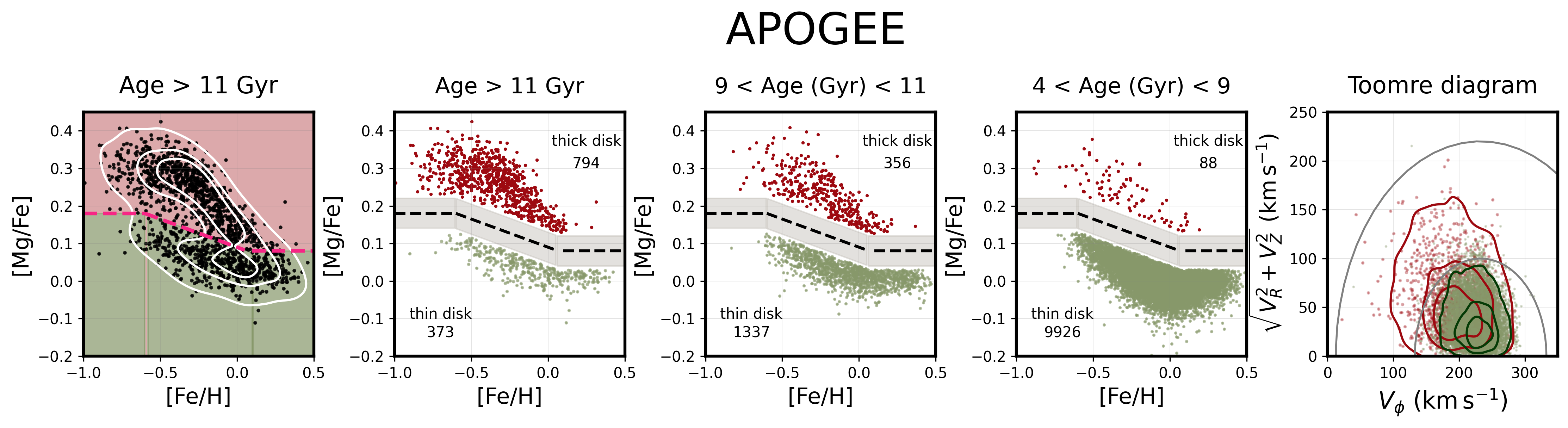}
    \includegraphics[width=2.1\columnwidth]{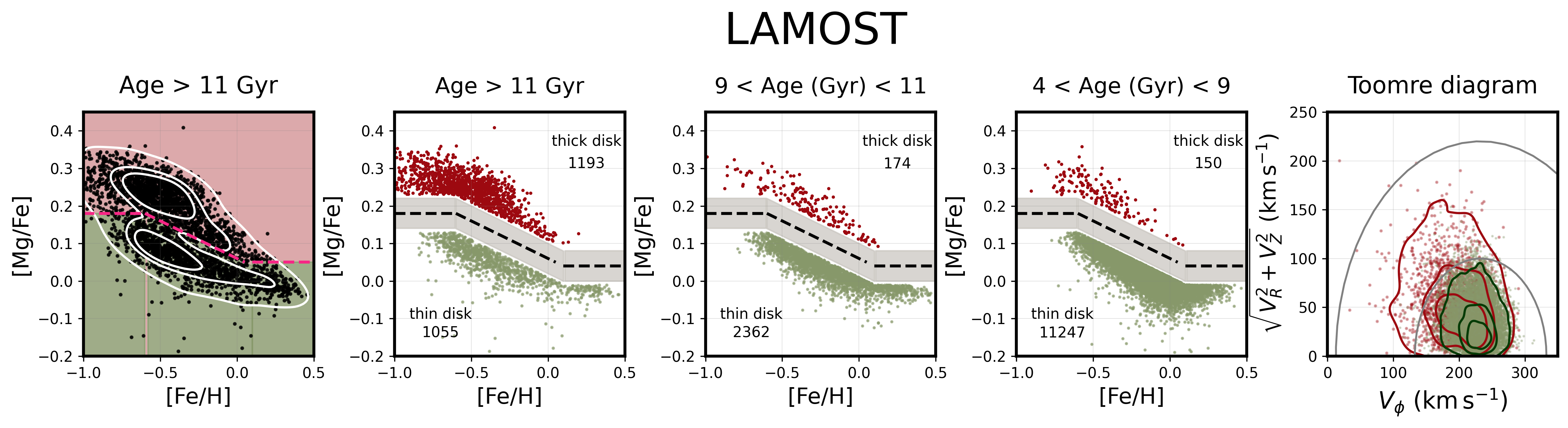}
    \caption{Distribution of APOGEE and LAMOST data in top and bottom rows, respectively. The panels from the first to the fourth column show the [Mg/Fe]--[Fe/H] space separated into different age intervals. The first column shows the density contours of data with ages older than 11\,Gyr. These contours were used to divide the chemical thick disk (red region, high-$\alpha$) and chemical thin disk (green region, low-$\alpha$) populations, the exact division being represented by the pink dashed line. For the APOGEE panel, the separation line represents $\rm[Mg/Fe] = 0.18$ for the interval $-1.0 < \rm[Fe/H] < -0.6$, $\rm[Mg/Fe] = -0.15 \times \rm[Fe/H] + 0.09$ between $-0.6 < \rm[Fe/H] < 0.1$ and $\rm[Mg/Fe] = 0.08$ for $\rm[Fe/H] > 0.1$. For the LAMOST panels, the curve is represented by $\rm[Mg/Fe] = 0.18$ for the interval $-1.0 < \rm[Fe/H] < -0.6$, $\rm[Mg/Fe] = -0.20 \times \rm[Fe/H] + 0.06$ between $-0.6 < \rm[Fe/H] < 0.1$ and $\rm[Mg/Fe] = 0.04$ for $\rm[Fe/H] > 0.1$. From the second to the fourth column, the data presented correspond to our final selection of thin (green dots) and thick disk (red dots) stars separated into three age ranges: older than 11\,Gyr, before the GSE accretion (11 - 9 Gyr ago), between 9\,Gyr and 11\,Gyr, merger period with the GSE, and between 4\,Gyr and 9\,Gyr, post-merger. The number of stars in each sample is also shown in the panels. Finally, the rightmost panel shows the projection of the thin and thick disk samples into Toomre diagram ($\sqrt{V_R^2 + V_Z^2} \text{ versus } V_{\phi}$ cylindrical frame), where the innermost curve represents $\rVert v - v_{circ} \rVert = 100$\,km\,s$^{-1}$ and the red curve is $\rVert v - v_{circ} \rVert = 180$\,km\,s$^{-1}$.}
    
    \label{data}
\end{figure*}
\subsection{APOGEE DR17}
\label{sec:APOGEE}

APOGEE contains observations with two high-resolution ($R \sim$ 22,000) spectrographs \citep{Wilson2019} installed on two 2.5m telescopes in the northern and southern hemispheres \citep{Bowen1973, Gunn2006}. The observations, made in the near-infrared (1.51$\mu$m--1.69$\mu$m; \textit{H}-band), provide stellar parameters and abundance of 15 chemical species for 150,000 stars across both hemispheres.

We apply some quality cuts to obtain stars with high-quality abundance measurements, excluding sources that exhibited any issues in the spectra, the spectral fitting process, and the estimated parameters (\texttt{STARFLAG} == 0; see \citealt{jonsson2020}). Additionally, we ensured accurate estimates of [Fe/H] and [Mg/Fe] by selecting sources with no flagged issues (i.e., flagged == 0). We also select stars with a good signal-to-noise ratio ($S/N > 50$)  which allows precise estimates of chemical abundances, radial velocities, and stellar parameters \citep{GarciaPerez2016, jonsson2020}.  After this selection, the APOGEE sample contains 20,026 stars with derived ages and age uncertainty $\sigma_{\text{AGE}} \leq 1$\,Gyr. 

\subsection{LAMOST MRS DR7}
\label{sec:LAMOST}

%\joao{would it be too troublesome to crossmatch the stars with this DR8 VAC? \url{https://ui.adsabs.harvard.edu/abs/2022MNRAS.517.4875L/abstract}? They obtain Al and Mn abundances for LAMOST DR8}

LAMOST spectroscopic survey data covers the wavelength range of 370nm to 900nm. The telescope array is installed at the Xinglong Station of the National Astronomical Observatory of China and the stellar parameter catalogs are divided into low-resolution (LRS) and medium-resolution (MRS). The resolution from the spectra of LAMOST MRS DR7 is $R \approx$ 7500. \citet{Queiroz2023} applied the $\sigma_{T_{\text{eff}}} < 300$\,K, $\sigma_{\text{log }g} < 0.5$ and $\sigma_{[Fe/H]} < 0.3$ as cuts in stellar parameter uncertainty. We added both signal-to-noise ratio ($S/N > 50$), and the effective temperature restriction (4000 $< T_{\text{eff}}$\,(K) $<$ 6500). After this selection, the LAMOST sample contains 34,263 stars with derived ages and $\sigma_{\text{AGE}} \leq 1$\,Gyr.

\subsection{Orbital Parameters}
\label{sec:orb}

We employed \textit{Gaia} DR3 data for on-sky positions and proper motions, combined with \texttt{StarHorse} heliocentric distances and APOGEE/LAMOST radial velocities, to calculate the 6D phase-space vector in a Galactocentric Cartesian frame. The position of the Sun relative to the Galactic Center was assumed to be $(X, Y, Z)_{\rm GC} = (-8.2, 0.0, 0.0)$ kpc \citep{BlandHawthorn2016}, with the local circular velocity $v_c = 232.8$\kms \citep{mcmillan2017}, and the solar motion with respect to the local standard of rest $(U_{\odot}, V_{\odot}, W_{\odot}) = (11.10, 12.24, 7.25)$\kms \citep{schonrich2010}.

Orbital calculations were performed using the \texttt{AGAMA} \citep{Vasiliev2019} library for 20 Gyr forward, adopting the Galactic potential model described by \citet{mcmillan2017}. To account for uncertainties, we performed 50 Monte Carlo realizations of each stellar orbit according to Gaussian distributions of its uncertainties in the phase-space coordinates. The kinematic and dynamical parameters were derived as the medians of the resulting distributions, with uncertainties represented by the $16^{\rm th}$ and $84^{\rm th}$ percentiles of the distributions.

\section{Thin and thick disk selection}
\label{sec:selection}

To study in detail the contribution attributed to the thin and thick disks individually and as a whole in the evolutionary history of our Galaxy, we selected our thin and thick disk samples using the [Mg/Fe]--[Fe/H] diagram. The cuts used to select the samples are represented in the first column of Figure \ref{data} that shows the projection of stars older than 11\,Gyr in the [Mg/Fe]--[Fe/H] space. The white contours show the density regions, which were used to trace the separation between the regions associated with the chemical thin disk (lower green area) and the chemical thick disk (upper red area). The exact separation is represented by the pink dashed line in the first column and by the black dashed line from the second to the fourth columns. Although this separation is based on this older age range ($>$\,11\,Gyr), it is valid for all ages. We excluded data within a 0.1 dex wide band centered around the black dashed line to reduce contamination, represented by the gray band in the panels from the second to the fourth column. Note that the exact boundary line is different for APOGEE and LAMOST data simply because of the different abundance scales of each survey. The APOGEE (LAMOST) thick and thin disk final samples contain 1,258 (1,790) and 16,223 (18,991) stars, respectively.

The final thick and thin disk samples are shown from the second to the fourth panel of Figure \ref{data}, where the data are projected on the [Mg/Fe]--[Fe/H] diagram split into three age intervals: (I) older than 11\,Gyr, representing stars that formed before the merger with GSE, (II) between 9\,Gyr and 11\,Gyr, during the GSE merger (\citealt{Gallart2019, Bonaca2020, Naidu2021, Montalban2021, Giribaldi2023}), and (III) between 4\,Gyr and 9\,Gyr, corresponding to the period following the GSE accretion. The rightmost panel of Figure \ref{data} shows the thin (green) and thick (red) disk samples, with no restriction on age range, projected onto the Toomre diagram, $\sqrt{V_R^2 + V_Z^2} \text{ versus } V_{\phi}$ cylindrical frame. The innermost curve represents $\rVert v - v_{circ} \rVert = 100$\,km\,s$^{-1}$, adopting $v_{circ} = 232.8$\,km\,s$^{-1}$ \citep{mcmillan2017}, that correspond to thin disk orbits, while the outermost curve is representative of the thick disk orbits. Stars beyond the outermost curve exhibit halo orbits. The projection of our samples on the Toomre diagram confirms that the separation in [Mg/Fe]-[Fe/H] is satisfactory for selection of the thin and thick disk populations. We tested our analysis by implementing a Toomre diagram kinematic constraint on the thin and thick disk sample selection and confirmed that the results did not change. For this reason, we chose to keep only the chemical criterion to not bias our kinematic parameter analyses.

%%%%%%%%%%%%%%%%%%%%%%%%%%%%%%%%%%%%%%%%%%
%%%%%%%%%%%% RESULTS %%%%%%%%%%%%%%%
%%%%%%%%%%%%%%%%%%%%%%%%%%%%%%%%%%%%%%%%%%

\section{Results}
\label{sec:results} 

\begin{figure}
    \centering
    \includegraphics[width=\columnwidth]{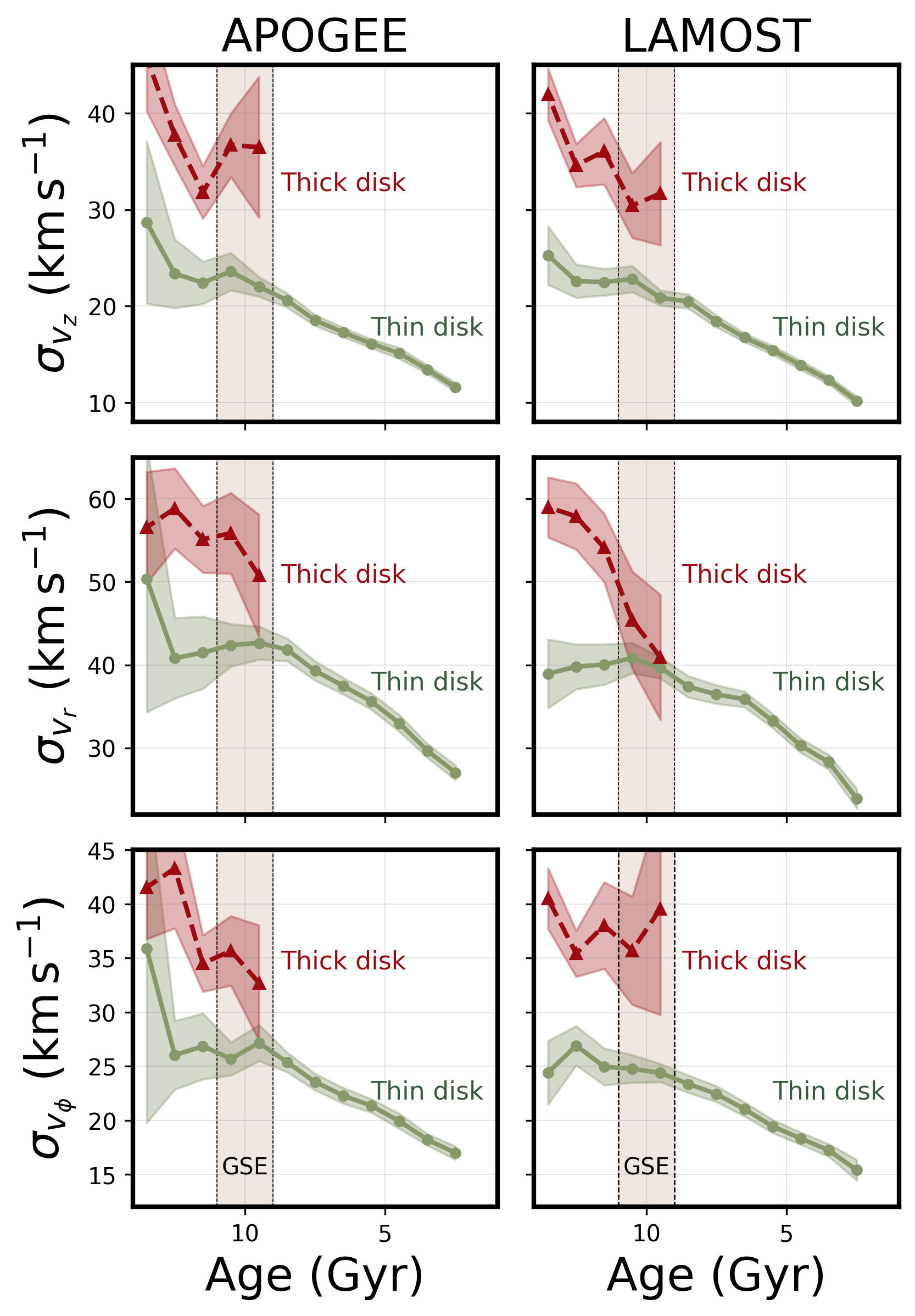}
    \caption{Velocity dispersion in $z$-direction ($\sigma_{V_z}$; upper panels), radial direction ($\sigma_{V_r}$; middle panels) and azimuthal direction ($\sigma_{V_\phi}$; bottom panels) as a function of \texttt{StarHorse} age for our thin (green line) and thick (red line) disk samples. The left and right columns show the APOGEE and LAMOST samples, respectively. The vertical band indicates the period of GSE merger (11 - 9 Gyr ago).}
    \label{std}
\end{figure}

The middle panels of Figure \ref{data} (age $> 11$\,Gyr) contain the oldest disk population born before the GSE accretion, that is composed mostly of stars classified as members of the thick disk in both surveys. Still, it includes a significant percentage of stars characteristic of the thin disk, being around $\sim$30\% of the full sample older than 11\,Gyr for APOGEE, and $\sim$45\% for LAMOST. It remains to be seen how much these exact fractions depend on specific target selection strategies by the different surveys. It is important to remember that all stars have uncertainties in age smaller than 1\,Gyr due to the selection chosen in this paper, which indicates that the stars in the middle panel are certainly older than 10\,Gyr at $\geq$1$\sigma$ level considering the maximum uncertainty in age. 

If there exists a genuinely old ($>$11\,Gyr) thin disk population, one would expect this population to be dynamically hotter than its young counterpart based on just secular evolution alone (\citealt{Nepal2024}). Figure \ref{std} shows the velocity dispersion in $z$- ($\sigma_{v_z}$), radial- ($\sigma_{v_r}$) and azimuthal ($\sigma_{v_\phi}$) directions as a function of \texttt{StarHorse} age for thin and thick disk populations. As expected, the thin disk samples have lower velocity dispersion than the thick disk for both APOGEE and LAMOST surveys independently of age. Furthermore, the old thin disk population exhibits higher $\sigma_{v_z}$, $\sigma_{v_r}$ and $\sigma_{v_\phi}$ compared to young stars in the same group. We observe a sharp drop in $\sigma_{v_z}$ from 10\,Gyr onwards, this decay is also observed in $\sigma_{v_r}$ and $\sigma_{v_\phi}$, and is coincident with the period just after the GSE merger, represented by the vertical band between 9--11\,Gyr. This behavior indicates that the population younger than the merger is kinematically cooler compared to the characteristics of the canonical thin disk population.

\begin{figure}
    \centering
    \includegraphics[width=\columnwidth]{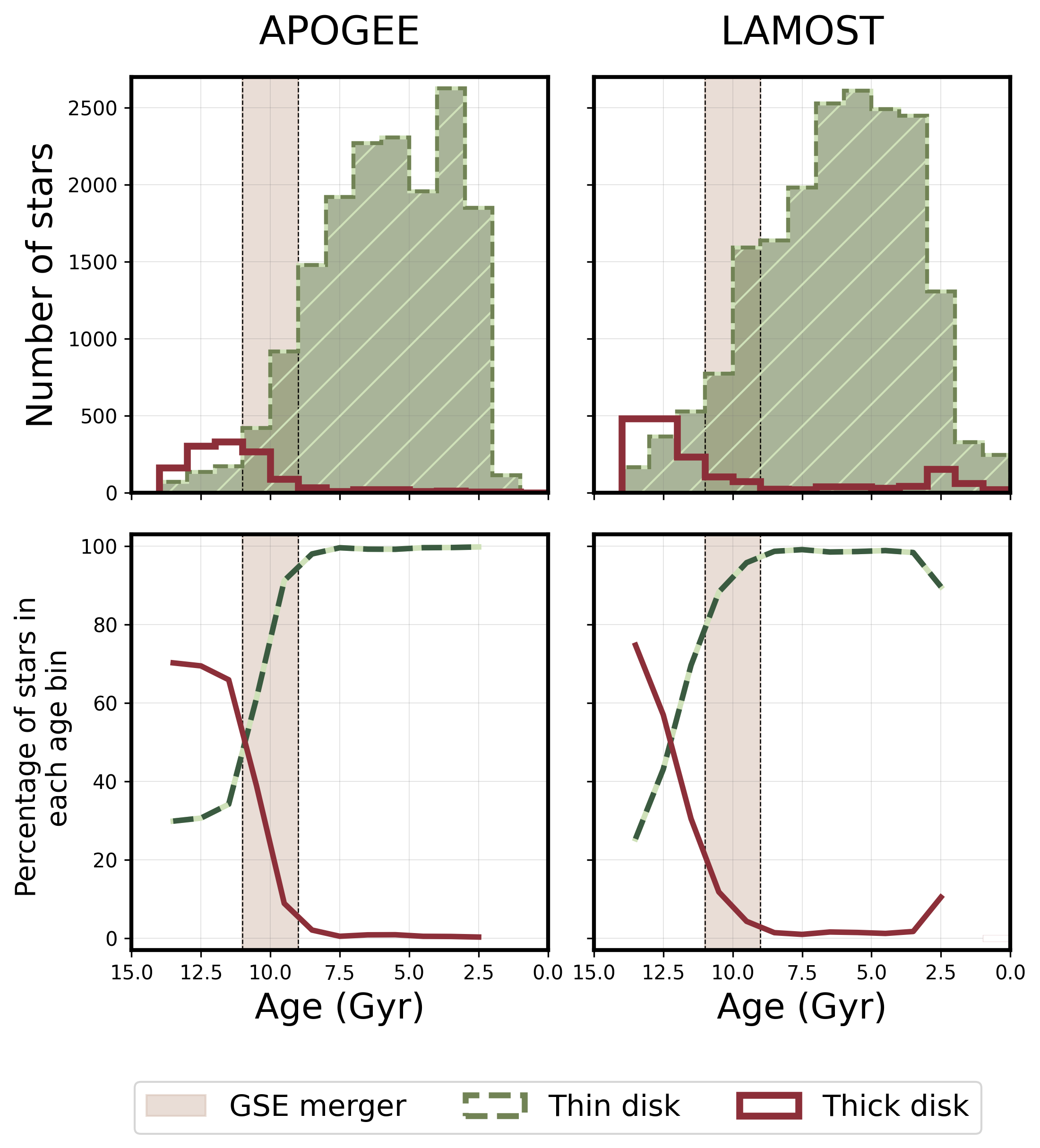}
    \caption{Distribution of \texttt{StarHorse} ages for APOGEE (top left) and LAMOST (top right panel) split into thin (green) and thick (red) disks. The bottom panels show the percentage of thin (green dashed) and thick (red solid line) disk stars in each age bin, considering age bins of 1\,Gyr. The sum of the thin and thick disks percentages totals 100\%. The vertical band indicates the period of GSE merger (9 -- 11 Gyr).} 
    \label{hist_new}
\end{figure}

\begin{figure*}[ht!]
    \centering
    \includegraphics[width=2.1\columnwidth]{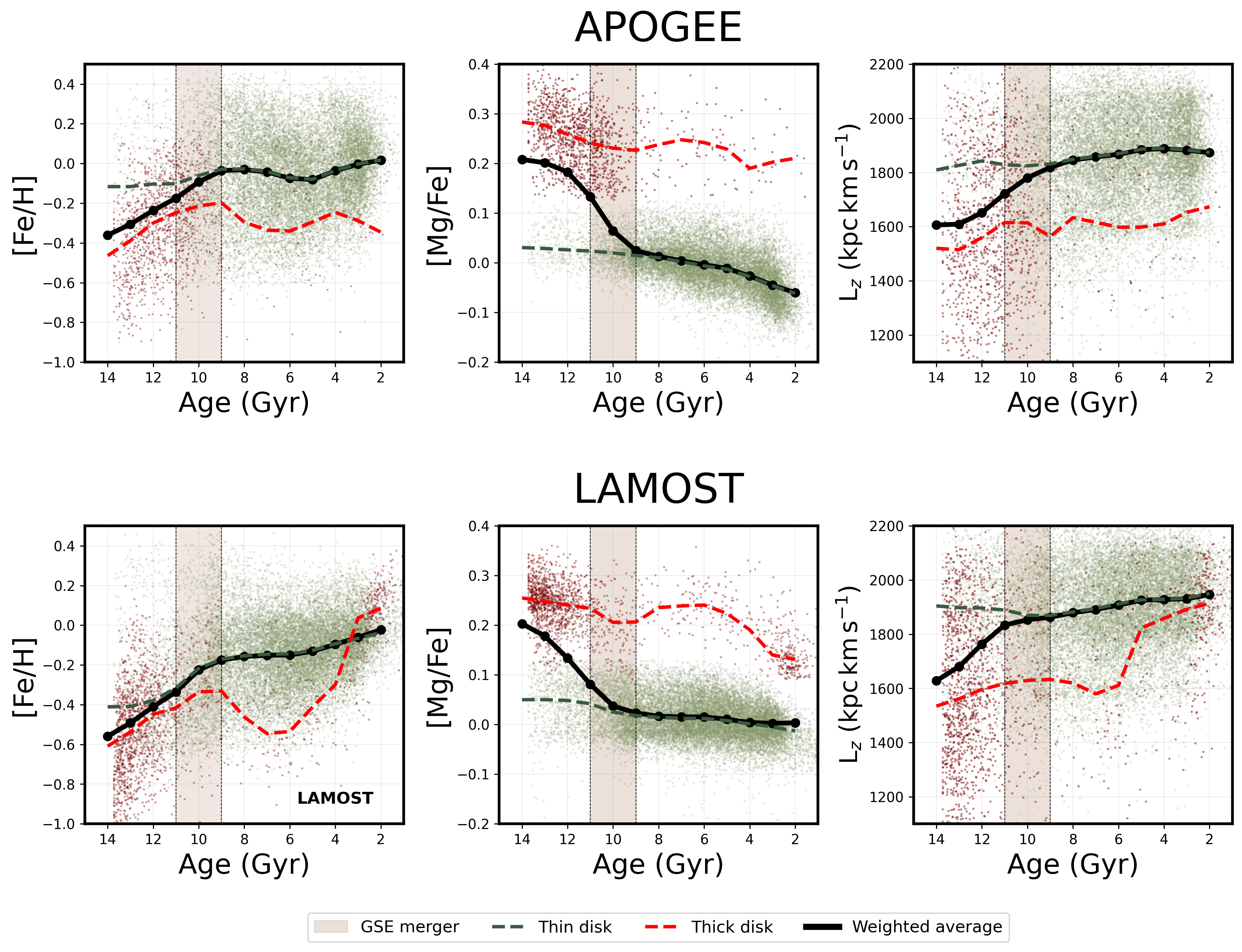}
    \caption{The columns were organized in the following sequence: [Fe/H] (first), [Mg/Fe] (second), and angular momentum in $z$-direction, $L_z$ (third) as a function of \texttt{StarHorse} age. The top and bottom rows show the APOGEE and LAMOST data, respectively. The thick disk is represented by red dots and the thin disk data by green dots. The black line represents the weighted average between the thick and thin disk samples, while the green and red lines represent the average for the thin and thick disk components, respectively. The vertical band indicates the period of GSE merger (11 - 9 Gyr).}
    \label{kde}
\end{figure*}

Figure \ref{hist_new} shows the \texttt{StarHorse} age distribution of thick and thin disk components in APOGEE (top left) and LAMOST (top right panel) samples. Almost the full thick disk sample is older than 10\,Gyr, resulting in a very old average age $\sim11.51\pm0.09$\,Gyr for APOGEE and $\sim12.19\pm0.11$\,Gyr for LAMOST, in which the age uncertainty was determined by bootstrapping method. The offset between the average ages found for the two surveys can be linked to the changes in the input parameters from one survey to the other, which may generate systematic differences. Only the APOGEE result is consistent with ages previously estimated in the literature (e.g., \citealt{Miglio2021, Queiroz2023}). The average age estimated for the thick disk by LAMOST is older than that found in \citet{Queiroz2023} for the same dataset (see Table 6 of their paper). This indicates that the value found in this paper appears to be an effect generated by sample selection. Note also that in \citet{Queiroz2023} the average ages of the thick disk in APOGEE DR17 and LAMOST MRS DR7 are consistent with each other. The presence of younger chemical thick disk stars is reported in previous studies (e.g., \citealt{Chiappini2015, Martig2015, Miglio2021, Queiroz2023, Grisoni2024}). Although the mechanism of origin of the young $\alpha$-rich is still under debate, there are some proposed interpretations. Most of the references quoted above suggest that these stars might be products of mass transfer or mergers, and therefore look younger (\citealt{Chiappini2015, Martig2015, Jofre2016, Izzard2018, Miglio2021, Jofre2023, Grisoni2024}). In this scenario, they should be present in all directions where the high-$\alpha$ population extends. However, these young $\alpha$-rich stars appeared to be more abundant towards the inner regions of the Galactic disk. For this reason, \citet{Chiappini2015} suggest a second interpretation in which the region of origin of these stars could be near the end of the Galactic bar, where the gas would remain inert for longer than in the spiral arms. The presence of young $\alpha$-rich stars has also been detected by other studies using different methods of age determination (e.g., \citealt{Haywood2013, Bensby2014, Bergemann2014}). The peak of $\alpha$-rich stars around 2\,Gyr in the bottom panel of Figure \ref{hist_new} appears to be a peculiarity in the LAMOST abundance estimates since it does not appear in the APOGEE data.

In contrast to the thick disk, the age distribution of the thin disk samples occupies a huge age range, ranging from the pre-merger period, showing a co-formation with the thick disk, to the present, as shown in the upper panels of Figure \ref{hist_new}. The number of stars in this component follows an increasing trend with age, with the mean of the distribution being $5.68\pm0.06$\,Gyr in APOGEE and $6.03\pm0.05$\,Gyr in LAMOST sample, within the range found by \citet{Queiroz2023}. The lower panels of Figure \ref{hist_new} show the percentage of stars classified as thin and thick disk in each age bin (considering 1\,Gyr bins), with the sum of these percentages being equal to 100\%. For our sample of SGB and MSTO stars of the APOGEE and LAMOST surveys, the thin disk population accounts for approximately 30\% of the total population at older ages ($\sim$ 13\,Gyr). The time at which each population reaches 50\% of the sample (intersection of the solid and dashed lines in the bottom panels of Figure \ref{hist_new}) is about 1 Gyr offset between surveys, expected due to systematic age uncertainties, but occurring at $\gtrsim$10\,Gyr ago in both samples. This difference may be due to systematics between stellar parameters from each survey being propagated to the \texttt{StarHorse} ages. Furthermore, it is important to highlight that solar neighborhood sample selection and the targeting of each survey can influence these percentages.

With stellar ages for our large sample of stars thanks to \texttt{StarHorse}, we can now provide a time-resolved picture of the MW's disk chemical and kinematic evolution. Figure \ref{kde} shows the behavior of [Fe/H], [Mg/Fe], and angular momentum in the $z$-direction ($L_{z}$) as a function of age. In this paper, we adopt $L_{z} > 0$ for stars with prograde orbits. The APOGEE and LAMOST data are present in the first and second rows, respectively. To analyze the fractions of thin and thick disks in the above-mentioned parameters, we calculated weighted averages of the contributions of each disk component as a function of age, considering bins of 1\,Gyr, which are represented by the black lines in all panels of Figure \ref{kde}. 

Figure \ref{kde} shows that in the period before the GSE accretion ($>$ 11 Gyr), our Galaxy is dominated by stars with chemical and kinematic properties of the thick disk. The trend is different from the post-GSE period ($<$ 9\,Gyr). The weighted average of $\rm[Fe/H]$ abundances (first columns) show a tendency to increase over time, but exhibit an approximately constant average value of $\rm[Fe/H] \sim -0.1$ at 9\,Gyr onward. This trend is accompanied by an opposite behavior for [Mg/Fe], which decreases from a higher pre-GSE value ($\rm[Mg/Fe] \sim +0.20$) to a lower post-GSE value ($\rm[Mg/Fe] \sim +0.05$). This indicates a clear transition from thick disk to thin disk dominance. The increase in $L_{z}$ (third column in Figure \ref{kde}) is kinematic evidence that corroborates with this scenario. Note that the thick disk contribution decreases significantly in the post-GSE period in all panels in Figure \ref{kde}. In addition, there is a very young peculiar population ($<$3\,Gyr) classified as thick disk in the LAMOST sample that must be a product of some variation in the LAMOST abundance estimates for stars in this age range, as mentioned above. However, this population does not affect our results, as we are interested in the early times of the MW. We also checked the amount of stellar rotation versus velocity, $V_{\phi}/\sigma_{\text{tot}}$ where  $\sigma_{\text{tot}} = \sqrt{\sigma_{V_r}^2 + \sigma_{V_z}^2 + \sigma_{V_{\phi}}^2}$, which is a good measure of disk rotation, and found that the thin disk has $V_{\phi}/\sigma_{\text{tot}} > 4$ for the entire age range, while the thick disk has $V_{\phi}/\sigma_{\text{tot}} > 2$. See Appendix \ref{sec:appendix_B} for more details.

%%%%%%%%%%%%%%%%%%%%%%%%%%%%%%%%%%%%%%%%%%
%%%%%%%%%%%% DISCUSSION %%%%%%%%%%%%%%%
%%%%%%%%%%%%%%%%%%%%%%%%%%%%%%%%%%%%%%%%%%

\section{Discussion}
\label{sec:discussion} 

The thin disk stars older than 11 Gyr were born before the GSE merger, unless this event happened significantly earlier than literature estimates. However, this possibility is unlikely because the dating of the merger is based on age estimates of red-giant stars with asteroseismology \citep{Montalban2021}, isochrone fitting \citep{Bonaca2020}, and globular clusters \citep{limberg2022}. Furthermore, the star-formation history of the GSE with photometric data and \textit{Gaia} astrometry \citep{Gallart2019, Gallart2024} and chemical evolution models for the GSE reinforce that the merger must have occurred around $z \sim 2$ \citep{Vincenzo2019, Hasselquist2021}. We also find that the velocity dispersion of the pre-GSE thin disk population is more significant than that of the young thin disk. This confirms the expectation that pre-GSE thin disk stars should have been born in a more dynamically heated environment at the beginning of the MW.
 
Our results show that the $\alpha$-element dichotomy between thin and thick disks was already in place since $>$\,11\,Gyr ago, which is not in agreement with models of sequential formation from the thick disk to the thin disk. Additionally, this difference in chemical enrichment may indicate that these populations were born in different environments simultaneously, where the star formation efficiency of the environment responsible for the thick disk would have been higher than the thin disk one. In this scenario, a higher star formation efficiency leads to more CCSNe events, which enrich the gas of the interstellar medium predominantly with $\alpha$-elements on a short timescale (e.g., \citealt{Chiappini2009IAU, Sheffield2012, Andrews2017, Grisoni2017, Grisoni2021}) and would explain the higher [Mg/Fe] abundance of the thick disk compared to the thin disk. 

Previous studies have reported the existence of this exquisitely old thin disk population with smaller samples and/or no detailed abundance data. \citet{Prudil2020} found 22 RR Lyrae stars in the solar neighborhood with $-1 < \rm[Fe/H] < 0$, low-$\alpha$, low eccentricity and low-$Z_{GC}$, properties shared by typical younger stars in the thin disk. RR Lyrae stars with disk orbits have also been recently indicated by \citet{DOrazi2024}. \citet{Beraldo2021} also reported the presence of thin disk stars older than 10\,Gyr using a sample of APOGEE DR16 \citep{APOGEE_DR16} stars and ages from \citet{Sanders&Das2018}. They argue that clump instabilities in the early disk offer a compelling explanation for the co-formation of thin and thick disks at early times. Most recently, \citet{Nepal2024} argued for the existence of these very old thin disk stars using data from \textit{Gaia} DR3 RVS spectra \citep{Cropper2018GaiaDR2rvs} and ages derived with \texttt{StarHorse}. In the same study, they show a population of old ($>$9\,Gyr) low-$\alpha$ stars that occupy the region in $V_{\phi} \text{ vs. }$[Fe/H] space expected for stars that were dynamically heated during the merger with GSE (see their Figure 7), known as the splash population \citep{DiMatteo2019, Belokurov2020}. Under the scenario that the merger with GSE triggers the formation of the thin disk, it is expected that only stars in the thick disk would have been heated. However, this low-$\alpha$ splash population reported in \citet{Nepal2024} is further evidence for the formation of the pre-merger thin disk. 

There is accumulating evidence for the scenario in which the thin disk begins its formation very early in the MW. This observational evidence brings new challenges to the study of the formation of the structure of our Galaxy, where this early ``co-formation'' of the MW disks puts into question proposed scenarios in the literature where a gas-rich merger is necessary to produce the observed low-/high-$\alpha$ dichotomy. This result also challenges scenarios where thick and thin disks are formed sequentially either mediated by a merger (e.g., \citealt{Grand2020, Buck2020alone}) or not (\citealt{Beane2024}).

Several previous simulation studies sought to reproduce the formation of the thin disk in MW-analogs from a gas infall (e.g., \citealt{Agertz2021, Renaud2021}). For example, \citet{Bignone2019} studied a MW-analog galaxy with similar merger history to analyze the impact of a GSE-analog merger into the galaxy disk. Their results indicate that the thin disk star formation rate (SFR) increases during the GSE-analog accretion. Also, \citet{Buck2020alone} used NIHAO-UHD simulations \citep{Buck2020a} to show that the low-$\alpha$ sequence (here associated with the thin disk) is generated from a gas-rich merger, which resets [Fe/H] and, with the decay of [$\alpha$/Fe] by SNeIa, produces the low-$\alpha$ population. These previous studies are based on a concept that the thin disk arises post-major merger. However, the observed pre-GSE thin disk stars in the MW raise a new question: \textit{how do these stars form, without this gas-rich merger, in the same period as the thick disk?}

Our results indicate that the GSE merger event does not trigger the thin disk formation, as previously suggested. However, we found that the period of this major merger is coincident with a significant increase in the number of thin disk stars, alongside a near truncation of star formation in the thick disk (see e.g. \citealt{Gallart2019, Bonaca2020}). 
 
While the period prior to the GSE merger is dominated by thick disk stars, the age interval after the GSE is populated mainly by thin disk stars. This leads us to interpret the GSE as an important mechanism in the transition between a predominant thick disk to a predominant thin disk, i.e., GSE may be essentially the reason for the quenching of the thick disk. The number of thin disk stars born after the GSE accretion is much larger than the number before this merger event in our samples (as shown in Figure \ref{data}), indicating a significant increase in this population. Although \citet{Gallart2024} do not indicate a thin disk population older than the GSE merger, they do show a peak in the history of thin disk star formation around the exactly time as this major merger (see figure 9 of that paper), which supports our result that the GSE can contribute with more gas to the formation of thin disk stars. In the future, tailored hydrodynamical simulations of MW-like galaxies that develop star-forming clumps with and without a GSE-like merger \citep[e.g.,][]{Amarante2022} could confirm (or not) our conjecture that, despite merger activity not being necessary for creating the low-/high-$\alpha$ dichotomy, it might be crucial in truncating the thick-disk formation, while accelerating the assembly of a chemical thin disk.

\section{Conclusions}
\label{conclusion} 

One of the most intriguing open questions in Galactic Archaeology is how the present-day thin and thick disk configuration in our Galaxy, the MW, came to be. Since these disks exhibit kinematic, spatial, and chemical differences, it is not trivial how these two populations were born and came to occupy their place in the MW disk. To contribute to the search for this answer, in this paper, we focus on the old disk population (age $> 11$\,Gyr) formed prior to the MW's last major merger with GSE. For this task, we combined chemical information from spectroscopic surveys (APOGEE DR17 and LAMOST MRS DR7), ages provided by the \texttt{StarHorse} code \citep{Queiroz2023} and astrometric data from \textit{Gaia} DR3 for MSTO and SGB stars. Our thin/thick disk samples are carefully based on the [Mg/Fe]--[Fe/H] space. 

Our main conclusions are summarized as follows:

\begin{itemize}

    \item [(i)] The period older than 11\,Gyr ago was populated by mostly high-$\alpha$, dynamically hotter, thick disk stars ($\sigma_{V_z} >$\,30\,km\,s$^{-1}$). A smaller fraction of stars in this period are classified with thin disk properties, low-$\alpha$, and dynamically colder ($\sigma_{V_z} <$\,20\,km\,s$^{-1}$) than the thick disk.

    \item [(ii)] We identified this ancient thin disk population ($> 11$\,Gyr) in both APOGEE and LAMOST surveys, existing in the same period associated with the formation of the thick disk, and before the GSE accretion. This result indicates a co-formation between the thick and thin disk at redshift $z > 2$, which puts into question scenarios in which the formation of the disks is sequential and major mergers are necessary to trigger the thin disk formation and $\alpha$-element bimodality. This very old thin disk population coincides with other old low-$\alpha$ populations mentioned in previous studies (e.g., \citealt{Laporte2020, Beraldo2021, Nepal2024}) and is also supported by the RR Lyrae stars with thin disk orbits and chemistry (e.g., \citealt{Prudil2020, Crestani2021, DOrazi2024}) and the low-$\alpha$ splash population reported in \citet{Nepal2024}. 

    \item[(iii)] Since the actual mechanism and location at which the thin and thick disk populations began to form are still unknown, it is difficult to measure the role of the GSE accretion in each of these thin and thick disk components. However, our analysis showed that the thin disk population grew significantly after this major merger, accompanied by a truncation of star formation in the thick disk. This indicates that the merger event with the GSE may mediate from an epoch dominated by the thick disk population to a Galaxy dominated by the thin disk.
\end{itemize}

Among the mechanisms proposed to explain the dichotomy between thin and thick disks, the formation of clumps in disk galaxies provides a favorable scenario for two populations with different chemical enrichments and orbital properties to co-form at the same period at early times in the MW. This, together with the contribution of the merger with the GSE, may have driven the conditions that led to the main features of the thin and thick disk configuration observed today.

\acknowledgements

L.B. thanks the partial financial support by the São Paulo Research Foundation (FAPESP), Brasil (Proc. 2024/16510-2), CAPES/PROEX (proc. 88887.821814/2023-00), and also thanks to all those involved with the multi-institutional \textit{Milky Way BR} Group for the weekly discussions. S.R. also thanks partial financial support from FAPESP (Proc.  2020/15245-2), CNPq (Proc. 303816/2022-8), and CAPES. H.D.P. thanks FAPESP (proc. 2024/17850-1). G.L. acknowledges support from KICP/UChicago through a KICP Postdoctoral Fellowship. JA is supported by the National Natural Science Foundation of China under grant No. 12233001, by the National Key R\&D Program of China under grant No. 2024YFA1611602, by a Shanghai Natural Science Research Grant (24ZR1491200), by the ``111'' project of the Ministry of Education under grant No. B20019, and by the China Manned Space Project with No. CMS-CSST-2021-B03. JA also thanks the sponsorship from Yangyang Development Fund. A.B.A acknowledge support from the Agencia Estatal de Investigación del Ministerio de Ciencia e Innovación (AEI-MCINN) under the grant "At the forefront of Galactic Archaeology: evolution of the luminous and dark matter
components of the Milky Way and Local Group dwarf galaxies in the Gaia era". This work was (partially) supported by the Spanish MICIN/AEI/10.13039/501100011033 and by "ERDF A way of making Europe" by the European Union through grant PID2021-122842OB-C21, and the Institute of Cosmos Sciences University of Barcelona (ICCUB, Unidad de Excelencia ’Mar\'{\i}a de Maeztu’) through grant CEX2019-000918-M. FA acknowledges the grant RYC2021-031683-I funded by MCIN/AEI/10.13039/501100011033 and by the European Union's NextGenerationEU/PRTR. F.O.B. thanks the financial support by the FAPESP (proc. 2022/16502-4). J. V. S. also thanks the financial support by the FAPESP (proc. 2024/22429-3). 

This work has made use of data from the European Space Agency (ESA) mission {\it Gaia} (\url{https://www.cosmos.esa.int/gaia}), processed by the {\it Gaia} Data Processing and Analysis Consortium (DPAC, \url{https://www.cosmos.esa.int/web/gaia/dpac/consortium}). Funding for the DPAC has been provided by national institutions, in particular, the institutions participating in the {\it Gaia} Multilateral Agreement.

\clearpage

\bibliographystyle{aasjournal}

\bibliography{Bibliography.bib}

\begin{thebibliography}{}
\expandafter\ifx\csname natexlab\endcsname\relax\def\natexlab#1{#1}\fi
\providecommand{\url}[1]{\href{#1}{#1}}
\providecommand{\dodoi}[1]{doi:~\href{http://doi.org/#1}{\nolinkurl{#1}}}
\providecommand{\doeprint}[1]{\href{http://ascl.net/#1}{\nolinkurl{http://ascl.net/#1}}}
\providecommand{\doarXiv}[1]{\href{https://arxiv.org/abs/#1}{\nolinkurl{https://arxiv.org/abs/#1}}}

\bibitem[{{Abdurro'uf} {et~al.}(2022){Abdurro'uf}, {Accetta}, {Aerts}, {Silva Aguirre}, {Ahumada}, {Ajgaonkar}, {Filiz Ak}, {Alam}, {Allende Prieto}, {Almeida}, {Anders}, {Anderson}, {Andrews}, {Anguiano}, {Aquino-Ort{\'\i}z}, {Arag{\'o}n-Salamanca}, {Argudo-Fern{\'a}ndez}, {Ata}, {Aubert}, {Avila-Reese}, {Badenes}, {Barb{\'a}}, {Barger}, {Barrera-Ballesteros}, {Beaton}, {Beers}, {Belfiore}, {Bender}, {Bernardi}, {Bershady}, {Beutler}, {Bidin}, {Bird}, {Bizyaev}, {Blanc}, {Blanton}, {Boardman}, {Bolton}, {Boquien}, {Borissova}, {Bovy}, {Brandt}, {Brown}, {Brownstein}, {Brusa}, {Buchner}, {Bundy}, {Burchett}, {Bureau}, {Burgasser}, {Cabang}, {Campbell}, {Cappellari}, {Carlberg}, {Wanderley}, {Carrera}, {Cash}, {Chen}, {Chen}, {Cherinka}, {Chiappini}, {Choi}, {Chojnowski}, {Chung}, {Clerc}, {Cohen}, {Comerford}, {Comparat}, {da Costa}, {Covey}, {Crane}, {Cruz-Gonzalez}, {Culhane}, {Cunha}, {Dai}, {Damke}, {Darling}, {Davidson}, {Davies}, {Dawson}, {De Lee}, {Diamond-Stanic}, {Cano-D{\'\i}az}, {S{\'a}nchez},
  {Donor}, {Duckworth}, {Dwelly}, {Eisenstein}, {Elsworth}, {Emsellem}, {Eracleous}, {Escoffier}, {Fan}, {Farr}, {Feng}, {Fern{\'a}ndez-Trincado}, {Feuillet}, {Filipp}, {Fillingham}, {Frinchaboy}, {Fromenteau}, {Galbany}, {Garc{\'\i}a}, {Garc{\'\i}a-Hern{\'a}ndez}, {Ge}, {Geisler}, {Gelfand}, {G{\'e}ron}, {Gibson}, {Goddy}, {Godoy-Rivera}, {Grabowski}, {Green}, {Greener}, {Grier}, {Griffith}, {Guo}, {Guy}, {Hadjara}, {Harding}, {Hasselquist}, {Hayes}, {Hearty}, {Hern{\'a}ndez}, {Hill}, {Hogg}, {Holtzman}, {Horta}, {Hsieh}, {Hsu}, {Hsu}, {Huber}, {Huertas-Company}, {Hutchinson}, {Hwang}, {Ibarra-Medel}, {Chitham}, {Ilha}, {Imig}, {Jaekle}, {Jayasinghe}, {Ji}, {Johnson}, {Jones}, {J{\"o}nsson}, {Katkov}, {Khalatyan}, {Kinemuchi}, {Kisku}, {Knapen}, {Kneib}, {Kollmeier}, {Kong}, {Kounkel}, {Kreckel}, {Krishnarao}, {Lacerna}, {Lane}, {Langgin}, {Lavender}, {Law}, {Lazarz}, {Leung}, {Leung}, {Lewis}, {Li}, {Li}, {Lian}, {Liang}, {Lin}, {Lin}, {Lin}, {Lintott}, {Long}, {Longa-Pe{\~n}a}, {L{\'o}pez-Cob{\'a}}, {Lu},
  {Lundgren}, {Luo}, {Mackereth}, {de la Macorra}, {Mahadevan}, {Majewski}, {Manchado}, {Mandeville}, {Maraston}, {Margalef-Bentabol}, {Masseron}, {Masters}, {Mathur}, {McDermid}, {Mckay}, {Merloni}, {Merrifield}, {Meszaros}, {Miglio}, {Di Mille}, {Minniti}, {Minsley}, \& {Monachesi}}]{Abdurrouf2022}
{Abdurro'uf}, {Accetta}, K., {Aerts}, C., {et~al.} 2022, \apjs, 259, 35, \dodoi{10.3847/1538-4365/ac4414}

\bibitem[{{Agertz} {et~al.}(2021){Agertz}, {Renaud}, {Feltzing}, {Read}, {Ryde}, {Andersson}, {Rey}, {Bensby}, \& {Feuillet}}]{Agertz2021}
{Agertz}, O., {Renaud}, F., {Feltzing}, S., {et~al.} 2021, \mnras, 503, 5826, \dodoi{10.1093/mnras/stab322}

\bibitem[{{Ahumada} {et~al.}(2020){Ahumada}, {Prieto}, {Almeida}, {Anders}, {Anderson}, {Andrews}, {Anguiano}, {Arcodia}, {Armengaud}, {Aubert}, {Avila}, {Avila-Reese}, {Badenes}, {Balland}, {Barger}, {Barrera-Ballesteros}, {Basu}, {Bautista}, {Beaton}, {Beers}, {Benavides}, {Bender}, {Bernardi}, {Bershady}, {Beutler}, {Bidin}, {Bird}, {Bizyaev}, {Blanc}, {Blanton}, {Boquien}, {Borissova}, {Bovy}, {Brandt}, {Brinkmann}, {Brownstein}, {Bundy}, {Bureau}, {Burgasser}, {Burtin}, {Cano-D{\'\i}az}, {Capasso}, {Cappellari}, {Carrera}, {Chabanier}, {Chaplin}, {Chapman}, {Cherinka}, {Chiappini}, {Doohyun Choi}, {Chojnowski}, {Chung}, {Clerc}, {Coffey}, {Comerford}, {Comparat}, {da Costa}, {Cousinou}, {Covey}, {Crane}, {Cunha}, {Ilha}, {Dai}, {Damsted}, {Darling}, {Davidson}, {Davies}, {Dawson}, {De}, {de la Macorra}, {De Lee}, {Queiroz}, {Deconto Machado}, {de la Torre}, {Dell'Agli}, {du Mas des Bourboux}, {Diamond-Stanic}, {Dillon}, {Donor}, {Drory}, {Duckworth}, {Dwelly}, {Ebelke}, {Eftekharzadeh}, {Davis
  Eigenbrot}, {Elsworth}, {Eracleous}, {Erfanianfar}, {Escoffier}, {Fan}, {Farr}, {Fern{\'a}ndez-Trincado}, {Feuillet}, {Finoguenov}, {Fofie}, {Fraser-McKelvie}, {Frinchaboy}, {Fromenteau}, {Fu}, {Galbany}, {Garcia}, {Garc{\'\i}a-Hern{\'a}ndez}, {Oehmichen}, {Ge}, {Maia}, {Geisler}, {Gelfand}, {Goddy}, {Gonzalez-Perez}, {Grabowski}, {Green}, {Grier}, {Guo}, {Guy}, {Harding}, {Hasselquist}, {Hawken}, {Hayes}, {Hearty}, {Hekker}, {Hogg}, {Holtzman}, {Horta}, {Hou}, {Hsieh}, {Huber}, {Hunt}, {Chitham}, {Imig}, {Jaber}, {Angel}, {Johnson}, {Jones}, {J{\"o}nsson}, {Jullo}, {Kim}, {Kinemuchi}, {Kirkpatrick}, {Kite}, {Klaene}, {Kneib}, {Kollmeier}, {Kong}, {Kounkel}, {Krishnarao}, {Lacerna}, {Lan}, {Lane}, {Law}, {Le Goff}, {Leung}, {Lewis}, {Li}, {Lian}, {Lin}, {Long}, {Longa-Pe{\~n}a}, {Lundgren}, {Lyke}, {Ted Mackereth}, {MacLeod}, {Majewski}, {Manchado}, {Maraston}, {Martini}, {Masseron}, {Masters}, {Mathur}, {McDermid}, {Merloni}, {Merrifield}, {M{\'e}sz{\'a}ros}, {Miglio}, {Minniti}, {Minsley}, {Miyaji},
  {Mohammad}, {Mosser}, {Mueller}, {Muna}, {Mu{\~n}oz-Guti{\'e}rrez}, {Myers}, {Nadathur}, {Nair}, {Nandra}, {do Nascimento}, {Nevin}, {Newman}, {Nidever}, {Nitschelm}, {Noterdaeme}, {O'Connell}, {Olmstead}, {Oravetz}, {Oravetz}, {Osorio}, {Pace}, {Padilla}, {Palanque-Delabrouille}, {Palicio}, {Pan}, {Pan}, {Parker}, {Paviot}, {Peirani}, {Ram{\'r}ez}, {Penny}, {Percival}, {Perez-Fournon}, {P{\'e}rez-R{\`a}fols}, {Petitjean}, {Pieri}, {Pinsonneault}, {Poovelil}, {Povick}, {Prakash}, {Price-Whelan}, {Raddick}, {Raichoor}, {Ray}, {Rembold}, {Rezaie}, {Riffel}, {Riffel}, {Rix}, {Robin}, {Roman-Lopes}, {Rom{\'a}n-Z{\'u}{\~n}iga}, {Rose}, {Ross}, {Rossi}, {Rowlands}, {Rubin}, {Salvato}, {S{\'a}nchez}, {S{\'a}nchez-Menguiano}, {S{\'a}nchez-Gallego}, {Sayres}, {Schaefer}, {Schiavon}, {Schimoia}, {Schlafly}, {Schlegel}, {Schneider}, {Schultheis}, {Schwope}, {Seo}, {Serenelli}, {Shafieloo}, {Shamsi}, {Shao}, {Shen}, {Shetrone}, {Shirley}, {Aguirre}, {Simon}, {Skrutskie}, {Slosar}, {Smethurst}, {Sobeck}, {Sodi},
  {Souto}, {Stark}, {Stassun}, {Steinmetz}, {Stello}, {Stermer}, {Storchi-Bergmann}, {Streblyanska}, {Stringfellow}, {Stutz}, {Su{\'a}rez}, {Sun}, {Taghizadeh-Popp}, {Talbot}, {Tayar}, {Thakar}, {Theriault}, {Thomas}, {Thomas}, {Tinker}, {Tojeiro}, {Toledo}, {Tremonti}, {Troup}, {Tuttle}, {Unda-Sanzana}, {Valentini}, {Vargas-Gonz{\'a}lez}, {Vargas-Maga{\~n}a}, {V{\'a}zquez-Mata}, {Vivek}, {Wake}, {Wang}, {Weaver}, {Weijmans}, {Wild}, {Wilson}, {Wilson}, {Wolthuis}, {Wood-Vasey}, {Yan}, {Yang}, {Y{\`e}che}, {Zamora}, {Zarrouk}, {Zasowski}, {Zhang}, {Zhao}, {Zhao}, {Zheng}, {Zheng}, {Zhu}, \& {Zou}}]{APOGEE_DR16}
{Ahumada}, R., {Prieto}, C.~A., {Almeida}, A., {et~al.} 2020, \apjs, 249, 3, \dodoi{10.3847/1538-4365/ab929e}

\bibitem[{{Amarante} {et~al.}(2020){Amarante}, {Beraldo e Silva}, {Debattista}, \& {Smith}}]{Amarante2020}
{Amarante}, J. A.~S., {Beraldo e Silva}, L., {Debattista}, V.~P., \& {Smith}, M.~C. 2020, \apjl, 891, L30, \dodoi{10.3847/2041-8213/ab78a4}

\bibitem[{{Amarante} {et~al.}(2022){Amarante}, {Debattista}, {Beraldo e Silva}, {Laporte}, \& {Deg}}]{Amarante2022}
{Amarante}, J. A.~S., {Debattista}, V.~P., {Beraldo e Silva}, L., {Laporte}, C. F.~P., \& {Deg}, N. 2022, \apj, 937, 12, \dodoi{10.3847/1538-4357/ac8b0d}

\bibitem[{{Anders} {et~al.}(2018){Anders}, {Chiappini}, {Santiago}, {Matijevi{\v{c}}}, {Queiroz}, {Steinmetz}, \& {Guiglion}}]{Anders2018}
{Anders}, F., {Chiappini}, C., {Santiago}, B.~X., {et~al.} 2018, \aap, 619, A125, \dodoi{10.1051/0004-6361/201833099}

\bibitem[{{Anders} {et~al.}(2014){Anders}, {Chiappini}, {Santiago}, {Rocha-Pinto}, {Girardi}, {da Costa}, {Maia}, {Steinmetz}, {Minchev}, {Schultheis}, {Boeche}, {Miglio}, {Montalb{\'a}n}, {Schneider}, {Beers}, {Cunha}, {Allende Prieto}, {Balbinot}, {Bizyaev}, {Brauer}, {Brinkmann}, {Frinchaboy}, {Garc{\'\i}a P{\'e}rez}, {Hayden}, {Hearty}, {Holtzman}, {Johnson}, {Kinemuchi}, {Majewski}, {Malanushenko}, {Malanushenko}, {Nidever}, {O'Connell}, {Pan}, {Robin}, {Schiavon}, {Shetrone}, {Skrutskie}, {Smith}, {Stassun}, \& {Zasowski}}]{Anders2014}
---. 2014, \aap, 564, A115, \dodoi{10.1051/0004-6361/201323038}

\bibitem[{{Anders} {et~al.}(2017){Anders}, {Chiappini}, {Rodrigues}, {Miglio}, {Montalb{\'a}n}, {Mosser}, {Girardi}, {Valentini}, {Noels}, {Morel}, {Johnson}, {Schultheis}, {Baudin}, {de Assis Peralta}, {Hekker}, {Theme{\ss}l}, {Kallinger}, {Garc{\'\i}a}, {Mathur}, {Baglin}, {Santiago}, {Martig}, {Minchev}, {Steinmetz}, {da Costa}, {Maia}, {Allende Prieto}, {Cunha}, {Beers}, {Epstein}, {Garc{\'\i}a P{\'e}rez}, {Garc{\'\i}a-Hern{\'a}ndez}, {Harding}, {Holtzman}, {Majewski}, {M{\'e}sz{\'a}ros}, {Nidever}, {Pan}, {Pinsonneault}, {Schiavon}, {Schneider}, {Shetrone}, {Stassun}, {Zamora}, \& {Zasowski}}]{Anders2017}
{Anders}, F., {Chiappini}, C., {Rodrigues}, T.~S., {et~al.} 2017, \aap, 597, A30, \dodoi{10.1051/0004-6361/201527204}

\bibitem[{{Anders} {et~al.}(2019){Anders}, {Khalatyan}, {Chiappini}, {Queiroz}, {Santiago}, {Jordi}, {Girardi}, {Brown}, {Matijevi{\v{c}}}, {Monari}, {Cantat-Gaudin}, {Weiler}, {Khan}, {Miglio}, {Carrillo}, {Romero-G{\'o}mez}, {Minchev}, {de Jong}, {Antoja}, {Ramos}, {Steinmetz}, \& {Enke}}]{Anders2019}
{Anders}, F., {Khalatyan}, A., {Chiappini}, C., {et~al.} 2019, \aap, 628, A94, \dodoi{10.1051/0004-6361/201935765}

\bibitem[{{Anders} {et~al.}(2022){Anders}, {Khalatyan}, {Queiroz}, {Chiappini}, {Ard{\`e}vol}, {Casamiquela}, {Figueras}, {Jim{\'e}nez-Arranz}, {Jordi}, {Mongui{\'o}}, {Romero-G{\'o}mez}, {Altamirano}, {Antoja}, {Assaad}, {Cantat-Gaudin}, {Castro-Ginard}, {Enke}, {Girardi}, {Guiglion}, {Khan}, {Luri}, {Miglio}, {Minchev}, {Ramos}, {Santiago}, \& {Steinmetz}}]{Anders2022}
{Anders}, F., {Khalatyan}, A., {Queiroz}, A.~B.~A., {et~al.} 2022, \aap, 658, A91, \dodoi{10.1051/0004-6361/202142369}

\bibitem[{{Andrews} {et~al.}(2017){Andrews}, {Weinberg}, {Sch{\"o}nrich}, \& {Johnson}}]{Andrews2017}
{Andrews}, B.~H., {Weinberg}, D.~H., {Sch{\"o}nrich}, R., \& {Johnson}, J.~A. 2017, \apj, 835, 224, \dodoi{10.3847/1538-4357/835/2/224}

\bibitem[{{Angl{\'e}s-Alc{\'a}zar} {et~al.}(2017){Angl{\'e}s-Alc{\'a}zar}, {Faucher-Gigu{\`e}re}, {Kere{\v{s}}}, {Hopkins}, {Quataert}, \& {Murray}}]{AnglesAlcazar2017}
{Angl{\'e}s-Alc{\'a}zar}, D., {Faucher-Gigu{\`e}re}, C.-A., {Kere{\v{s}}}, D., {et~al.} 2017, \mnras, 470, 4698, \dodoi{10.1093/mnras/stx1517}

\bibitem[{{Beane} {et~al.}(2024){Beane}, {Johnson}, {Semenov}, {Hernquist}, {Chandra}, \& {Conroy}}]{Beane2024}
{Beane}, A., {Johnson}, J., {Semenov}, V., {et~al.} 2024, arXiv e-prints, arXiv:2410.21580, \dodoi{10.48550/arXiv.2410.21580}

\bibitem[{{Belokurov} {et~al.}(2018){Belokurov}, {Erkal}, {Evans}, {Koposov}, \& {Deason}}]{belokurov2018}
{Belokurov}, V., {Erkal}, D., {Evans}, N.~W., {Koposov}, S.~E., \& {Deason}, A.~J. 2018, \mnras, 478, 611, \dodoi{10.1093/mnras/sty982}

\bibitem[{{Belokurov} {et~al.}(2020){Belokurov}, {Sanders}, {Fattahi}, {Smith}, {Deason}, {Evans}, \& {Grand}}]{Belokurov2020}
{Belokurov}, V., {Sanders}, J.~L., {Fattahi}, A., {et~al.} 2020, \mnras, 494, 3880, \dodoi{10.1093/mnras/staa876}

\bibitem[{{Bensby} {et~al.}(2003){Bensby}, {Feltzing}, \& {Lundstr{\"o}m}}]{Bensby2003}
{Bensby}, T., {Feltzing}, S., \& {Lundstr{\"o}m}, I. 2003, \aap, 410, 527, \dodoi{10.1051/0004-6361:20031213}

\bibitem[{{Bensby} {et~al.}(2014){Bensby}, {Feltzing}, \& {Oey}}]{Bensby2014}
{Bensby}, T., {Feltzing}, S., \& {Oey}, M.~S. 2014, \aap, 562, A71, \dodoi{10.1051/0004-6361/201322631}

\bibitem[{{Beraldo e Silva} {et~al.}(2021){Beraldo e Silva}, {Debattista}, {Nidever}, {Amarante}, \& {Garver}}]{Beraldo2021}
{Beraldo e Silva}, L., {Debattista}, V.~P., {Nidever}, D., {Amarante}, J. A.~S., \& {Garver}, B. 2021, \mnras, 502, 260, \dodoi{10.1093/mnras/staa3966}

\bibitem[{{Bergemann} {et~al.}(2014){Bergemann}, {Ruchti}, {Serenelli}, {Feltzing}, {Alves-Brito}, {Asplund}, {Bensby}, {Gruyters}, {Heiter}, {Hourihane}, {Korn}, {Lind}, {Marino}, {Jofre}, {Nordlander}, {Ryde}, {Worley}, {Gilmore}, {Randich}, {Ferguson}, {Jeffries}, {Micela}, {Negueruela}, {Prusti}, {Rix}, {Vallenari}, {Alfaro}, {Allende Prieto}, {Bragaglia}, {Koposov}, {Lanzafame}, {Pancino}, {Recio-Blanco}, {Smiljanic}, {Walton}, {Costado}, {Franciosini}, {Hill}, {Lardo}, {de Laverny}, {Magrini}, {Maiorca}, {Masseron}, {Morbidelli}, {Sacco}, {Kordopatis}, \& {Tautvai{\v{s}}ien{\.{e}}}}]{Bergemann2014}
{Bergemann}, M., {Ruchti}, G.~R., {Serenelli}, A., {et~al.} 2014, \aap, 565, A89, \dodoi{10.1051/0004-6361/201423456}

\bibitem[{{Bignone} {et~al.}(2019){Bignone}, {Helmi}, \& {Tissera}}]{Bignone2019}
{Bignone}, L.~A., {Helmi}, A., \& {Tissera}, P.~B. 2019, \apjl, 883, L5, \dodoi{10.3847/2041-8213/ab3e0e}

\bibitem[{{Bland-Hawthorn} \& {Gerhard}(2016)}]{BlandHawthorn2016}
{Bland-Hawthorn}, J., \& {Gerhard}, O. 2016, \araa, 54, 529, \dodoi{10.1146/annurev-astro-081915-023441}

\bibitem[{{Bonaca} {et~al.}(2020){Bonaca}, {Conroy}, {Cargile}, {Naidu}, {Johnson}, {Zaritsky}, {Ting}, {Caldwell}, {Han}, \& {van Dokkum}}]{Bonaca2020}
{Bonaca}, A., {Conroy}, C., {Cargile}, P.~A., {et~al.} 2020, \apjl, 897, L18, \dodoi{10.3847/2041-8213/ab9caa}

\bibitem[{{Bond} {et~al.}(2010){Bond}, {Ivezi{\'c}}, {Sesar}, {Juri{\'c}}, {Munn}, {Kowalski}, {Loebman}, {Ro{\v{s}}kar}, {Beers}, {Dalcanton}, {Rockosi}, {Yanny}, {Newberg}, {Allende Prieto}, {Wilhelm}, {Lee}, {Sivarani}, {Majewski}, {Norris}, {Bailer-Jones}, {Re Fiorentin}, {Schlegel}, {Uomoto}, {Lupton}, {Knapp}, {Gunn}, {Covey}, {Allyn Smith}, {Miknaitis}, {Doi}, {Tanaka}, {Fukugita}, {Kent}, {Finkbeiner}, {Quinn}, {Hawley}, {Anderson}, {Kiuchi}, {Chen}, {Bushong}, {Sohi}, {Haggard}, {Kimball}, {McGurk}, {Barentine}, {Brewington}, {Harvanek}, {Kleinman}, {Krzesinski}, {Long}, {Nitta}, {Snedden}, {Lee}, {Pier}, {Harris}, {Brinkmann}, \& {Schneider}}]{Bond2010}
{Bond}, N.~A., {Ivezi{\'c}}, {\v{Z}}., {Sesar}, B., {et~al.} 2010, \apj, 716, 1, \dodoi{10.1088/0004-637X/716/1/1}

\bibitem[{{Bovy} {et~al.}(2012){Bovy}, {Rix}, {Liu}, {Hogg}, {Beers}, \& {Lee}}]{Bovy2012}
{Bovy}, J., {Rix}, H.-W., {Liu}, C., {et~al.} 2012, \apj, 753, 148, \dodoi{10.1088/0004-637X/753/2/148}

\bibitem[{{Bowen} \& {Vaughan}(1973)}]{Bowen1973}
{Bowen}, I.~S., \& {Vaughan}, A.~H., J. 1973, \ao, 12, 1430, \dodoi{10.1364/AO.12.001430}

\bibitem[{Bressan {et~al.}(2012)Bressan, Marigo, Girardi, Salasnich, Dal~Cero, Rubele, \& Nanni}]{Bressan2012}
Bressan, A., Marigo, P., Girardi, L., {et~al.} 2012, Monthly Notices of the Royal Astronomical Society, 427, 127, \dodoi{10.1111/j.1365-2966.2012.21948.x}

\bibitem[{{Buck}(2020)}]{Buck2020alone}
{Buck}, T. 2020, \mnras, 491, 5435, \dodoi{10.1093/mnras/stz3289}

\bibitem[{{Buck} {et~al.}(2020){Buck}, {Obreja}, {Macci{\`o}}, {Minchev}, {Dutton}, \& {Ostriker}}]{Buck2020a}
{Buck}, T., {Obreja}, A., {Macci{\`o}}, A.~V., {et~al.} 2020, \mnras, 491, 3461, \dodoi{10.1093/mnras/stz3241}

\bibitem[{{Burrows} \& {Vartanyan}(2021)}]{Burrows2021}
{Burrows}, A., \& {Vartanyan}, D. 2021, \nat, 589, 29, \dodoi{10.1038/s41586-020-03059-w}

\bibitem[{{Chambers} {et~al.}(2016){Chambers}, {Magnier}, {Metcalfe}, {Flewelling}, {Huber}, {Waters}, {Denneau}, {Draper}, {Farrow}, \& {Finkbeiner}}]{Chambers2016}
{Chambers}, K.~C., {Magnier}, E.~A., {Metcalfe}, N., {et~al.} 2016, ArXiv e-prints.
\newblock \url{http://adsabs.harvard.edu/abs/2016arXiv161205560C}

\bibitem[{{Chiappini}(2009)}]{Chiappini2009IAU}
{Chiappini}, C. 2009, in IAU Symposium, Vol. 254, The Galaxy Disk in Cosmological Context, ed. J.~{Andersen}, {Nordstr{\"o}ara}, B.~{m}, \& J.~{Bland-Hawthorn}, 191--196, \dodoi{10.1017/S1743921308027580}

\bibitem[{{Chiappini} {et~al.}(1997){Chiappini}, {Matteucci}, \& {Gratton}}]{Chiappini1997}
{Chiappini}, C., {Matteucci}, F., \& {Gratton}, R. 1997, \apj, 477, 765, \dodoi{10.1086/303726}

\bibitem[{{Chiappini} {et~al.}(2015){Chiappini}, {Anders}, {Rodrigues}, {Miglio}, {Montalb{\'a}n}, {Mosser}, {Girardi}, {Valentini}, {Noels}, {Morel}, {Minchev}, {Steinmetz}, {Santiago}, {Schultheis}, {Martig}, {da Costa}, {Maia}, {Allende Prieto}, {de Assis Peralta}, {Hekker}, {Theme{\ss}l}, {Kallinger}, {Garc{\'\i}a}, {Mathur}, {Baudin}, {Beers}, {Cunha}, {Harding}, {Holtzman}, {Majewski}, {M{\'e}sz{\'a}ros}, {Nidever}, {Pan}, {Schiavon}, {Shetrone}, {Schneider}, \& {Stassun}}]{Chiappini2015}
{Chiappini}, C., {Anders}, F., {Rodrigues}, T.~S., {et~al.} 2015, \aap, 576, L12, \dodoi{10.1051/0004-6361/201525865}

\bibitem[{{Clarke} {et~al.}(2019){Clarke}, {Debattista}, {Nidever}, {Loebman}, {Simons}, {Kassin}, {Du}, {Ness}, {Fisher}, {Quinn}, {Wadsley}, {Freeman}, \& {Popescu}}]{Clarke2019}
{Clarke}, A.~J., {Debattista}, V.~P., {Nidever}, D.~L., {et~al.} 2019, \mnras, 484, 3476, \dodoi{10.1093/mnras/stz104}

\bibitem[{{Crestani} {et~al.}(2021){Crestani}, {Braga}, {Fabrizio}, {Bono}, {Sneden}, {Preston}, {Ferraro}, {Iannicola}, {Nonino}, {Fiorentino}, {Th{\'e}venin}, {Lemasle}, {Prudil}, {Alves-Brito}, {Altavilla}, {Chaboyer}, {Dall'Ora}, {D'Orazi}, {Gilligan}, {Grebel}, {Koch-Hansen}, {Lala}, {Marengo}, {Marinoni}, {Marrese}, {Mart{\'\i}nez-V{\'a}zquez}, {Matsunaga}, {Monelli}, {Mullen}, {Neeley}, {da Silva}, {Stetson}, {Salaris}, {Storm}, {Valenti}, \& {Zoccali}}]{Crestani2021}
{Crestani}, J., {Braga}, V.~F., {Fabrizio}, M., {et~al.} 2021, \apj, 914, 10, \dodoi{10.3847/1538-4357/abfa23}

\bibitem[{{Cropper} {et~al.}(2018){Cropper}, {Katz}, {Sartoretti}, {Prusti}, {de Bruijne}, {Chassat}, {Charvet}, {Boyadjian}, {Perryman}, {Sarri}, {Gare}, {Erdmann}, {Munari}, {Zwitter}, {Wilkinson}, {Arenou}, {Vallenari}, {G{\'o}mez}, {Panuzzo}, {Seabroke}, {Allende Prieto}, {Benson}, {Marchal}, {Huckle}, {Smith}, {Dolding}, {Jan{\ss}en}, {Viala}, {Blomme}, {Baker}, {Boudreault}, {Crifo}, {Soubiran}, {Fr{\'e}mat}, {Jasniewicz}, {Guerrier}, {Guy}, {Turon}, {Jean-Antoine-Piccolo}, {Th{\'e}venin}, {David}, {Gosset}, \& {Damerdji}}]{Cropper2018GaiaDR2rvs}
{Cropper}, M., {Katz}, D., {Sartoretti}, P., {et~al.} 2018, \aap, 616, A5, \dodoi{10.1051/0004-6361/201832763}

\bibitem[{{Cui} {et~al.}(2012){Cui}, {Zhao}, {Chu}, {Li}, {Li}, {Zhang}, {Su}, {Yao}, {Wang}, {Xing}, {Li}, {Zhu}, {Wang}, {Gu}, {Luo}, {Xu}, {Zhang}, {Liu}, {Zhang}, {Yang}, {Cao}, {Chen}, {Chen}, {Chen}, {Chen}, {Chu}, {Feng}, {Gong}, {Hou}, {Hu}, {Hu}, {Hu}, {Jia}, {Jiang}, {Jiang}, {Jiang}, {Jin}, {Li}, {Li}, {Li}, {Liu}, {Liu}, {Lu}, {Mao}, {Men}, {Qi}, {Qi}, {Shi}, {Tang}, {Tao}, {Wang}, {Wang}, {Wang}, {Wang}, {Wang}, {Wang}, {Wang}, {Wang}, {Wang}, {Wang}, {Wang}, {Wang}, {Xu}, {Xu}, {Yang}, {Yu}, {Yuan}, {Yuan}, {Zhai}, {Zhang}, {Zhang}, {Zhang}, {Zhao}, {Zhou}, {Zhou}, {Zhu}, \& {Zou}}]{LAMOST2012b}
{Cui}, X.-Q., {Zhao}, Y.-H., {Chu}, Y.-Q., {et~al.} 2012, Research in Astronomy and Astrophysics, 12, 1197, \dodoi{10.1088/1674-4527/12/9/003}

\bibitem[{De~Silva {et~al.}(2015)De~Silva, Freeman, Bland-Hawthorn, Martell, De~Boer, Asplund, Keller, Sharma, Zucker, Zwitter, {et~al.}}]{de2015galah}
De~Silva, G.~M., Freeman, K.~C., Bland-Hawthorn, J., {et~al.} 2015, Monthly Notices of the Royal Astronomical Society, 449, 2604

\bibitem[{{Debattista} {et~al.}(2023){Debattista}, {Liddicott}, {Gonzalez}, {Beraldo e Silva}, {Amarante}, {Lazar}, {Zoccali}, {Valenti}, {Fisher}, {Khachaturyants}, {Nidever}, {Quinn}, {Du}, \& {Kassin}}]{Debattista2023}
{Debattista}, V.~P., {Liddicott}, D.~J., {Gonzalez}, O.~A., {et~al.} 2023, \apj, 946, 118, \dodoi{10.3847/1538-4357/acbb00}

\bibitem[{{Di Matteo} {et~al.}(2019){Di Matteo}, {Haywood}, {Lehnert}, {Katz}, {Khoperskov}, {Snaith}, {G{\'o}mez}, \& {Robichon}}]{DiMatteo2019}
{Di Matteo}, P., {Haywood}, M., {Lehnert}, M.~D., {et~al.} 2019, \aap, 632, A4, \dodoi{10.1051/0004-6361/201834929}

\bibitem[{{D'Orazi} {et~al.}(2024){D'Orazi}, {Storm}, {Casey}, {Braga}, {Zocchi}, {Bono}, {Fabrizio}, {Sneden}, {Massari}, {Giribaldi}, {Bergemann}, {Campbell}, {Casagrande}, {de Grijs}, {De Silva}, {Lugaro}, {Zucker}, {Bragaglia}, {Feuillet}, {Fiorentino}, {Chaboyer}, {Dall'Ora}, {Marengo}, {Mart{\'\i}nez-V{\'a}zquez}, {Matsunaga}, {Monelli}, {Mullen}, {Nataf}, {Tantalo}, {Thevenin}, {Vitello}, {Kudritzki}, {Bland-Hawthorn}, {Buder}, {Freeman}, {Kos}, {Lewis}, {Lind}, {Martell}, {Sharma}, {Stello}, \& {Zwitter}}]{DOrazi2024}
{D'Orazi}, V., {Storm}, N., {Casey}, A.~R., {et~al.} 2024, \mnras, 531, 137, \dodoi{10.1093/mnras/stae1149}

\bibitem[{{Freeman} \& {Bland-Hawthorn}(2002)}]{Freeman2002}
{Freeman}, K., \& {Bland-Hawthorn}, J. 2002, \araa, 40, 487, \dodoi{10.1146/annurev.astro.40.060401.093840}

\bibitem[{{Fuhrmann}(1998)}]{Fuhrmann1998}
{Fuhrmann}, K. 1998, \aap, 338, 161

\bibitem[{{Fuhrmann}(2011)}]{Fuhrmann2011}
---. 2011, \mnras, 414, 2893, \dodoi{10.1111/j.1365-2966.2011.18476.x}

\bibitem[{{Gaia Collaboration} {et~al.}(2016){Gaia Collaboration}, {Prusti}, {de Bruijne}, {Brown}, {Vallenari}, {Babusiaux}, {Bailer-Jones}, {Bastian}, {Biermann}, {Evans}, {Eyer}, {Jansen}, {Jordi}, {Klioner}, {Lammers}, {Lindegren}, {Luri}, {Mignard}, {Milligan}, {Panem}, {Poinsignon}, {Pourbaix}, {Randich}, {Sarri}, {Sartoretti}, {Siddiqui}, {Soubiran}, {Valette}, {van Leeuwen}, {Walton}, {Aerts}, {Arenou}, {Cropper}, {Drimmel}, {H{\o}g}, {Katz}, {Lattanzi}, {O'Mullane}, {Grebel}, {Holland}, {Huc}, {Passot}, {Bramante}, {Cacciari}, {Casta{\~n}eda}, {Chaoul}, {Cheek}, {De Angeli}, {Fabricius}, {Guerra}, {Hern{\'a}ndez}, {Jean-Antoine-Piccolo}, {Masana}, {Messineo}, {Mowlavi}, {Nienartowicz}, {Ord{\'o}{\~n}ez-Blanco}, {Panuzzo}, {Portell}, {Richards}, {Riello}, {Seabroke}, {Tanga}, {Th{\'e}venin}, {Torra}, {Els}, {Gracia-Abril}, {Comoretto}, {Garcia-Reinaldos}, {Lock}, {Mercier}, {Altmann}, {Andrae}, {Astraatmadja}, {Bellas-Velidis}, {Benson}, {Berthier}, {Blomme}, {Busso}, {Carry}, {Cellino}, {Clementini},
  {Cowell}, {Creevey}, {Cuypers}, {Davidson}, {De Ridder}, {de Torres}, {Delchambre}, {Dell'Oro}, {Ducourant}, {Fr{\'e}mat}, {Garc{\'\i}a-Torres}, {Gosset}, {Halbwachs}, {Hambly}, {Harrison}, {Hauser}, {Hestroffer}, {Hodgkin}, {Huckle}, {Hutton}, {Jasniewicz}, {Jordan}, {Kontizas}, {Korn}, {Lanzafame}, {Manteiga}, {Moitinho}, {Muinonen}, {Osinde}, {Pancino}, {Pauwels}, {Petit}, {Recio-Blanco}, {Robin}, {Sarro}, {Siopis}, {Smith}, {Smith}, {Sozzetti}, {Thuillot}, {van Reeven}, {Viala}, {Abbas}, {Abreu Aramburu}, {Accart}, {Aguado}, {Allan}, {Allasia}, {Altavilla}, {{\'A}lvarez}, {Alves}, {Anderson}, {Andrei}, {Anglada Varela}, {Antiche}, {Antoja}, {Ant{\'o}n}, {Arcay}, {Atzei}, {Ayache}, {Bach}, {Baker}, {Balaguer-N{\'u}{\~n}ez}, {Barache}, {Barata}, {Barbier}, {Barblan}, {Baroni}, {Barrado y Navascu{\'e}s}, {Barros}, {Barstow}, {Becciani}, {Bellazzini}, {Bellei}, {Bello Garc{\'\i}a}, {Belokurov}, {Bendjoya}, {Berihuete}, {Bianchi}, {Bienaym{\'e}}, {Billebaud}, {Blagorodnova}, {Blanco-Cuaresma}, {Boch},
  {Bombrun}, {Borrachero}, {Bouquillon}, {Bourda}, {Bouy}, {Bragaglia}, {Breddels}, {Brouillet}, {Br{\"u}semeister}, {Bucciarelli}, {Budnik}, {Burgess}, {Burgon}, {Burlacu}, {Busonero}, {Buzzi}, {Caffau}, {Cambras}, {Campbell}, {Cancelliere}, {Cantat-Gaudin}, {Carlucci}, {Carrasco}, {Castellani}, {Charlot}, {Charnas}, {Charvet}, {Chassat}, {Chiavassa}, {Clotet}, {Cocozza}, {Collins}, {Collins}, {Costigan}, {Crifo}, {Cross}, {Crosta}, {Crowley}, {Dafonte}, {Damerdji}, {Dapergolas}, {David}, {David}, {De Cat}, {de Felice}, {de Laverny}, {De Luise}, {De March}, {de Martino}, {de Souza}, {Debosscher}, {del Pozo}, {Delbo}, {Delgado}, {Delgado}, {di Marco}, {Di Matteo}, {Diakite}, {Distefano}, {Dolding}, {Dos Anjos}, {Drazinos}, {Dur{\'a}n}, {Dzigan}, {Ecale}, {Edvardsson}, {Enke}, {Erdmann}, {Escolar}, {Espina}, {Evans}, {Eynard Bontemps}, {Fabre}, {Fabrizio}, {Faigler}, {Falc{\~a}o}, {Farr{\`a}s Casas}, {Faye}, {Federici}, {Fedorets}, {Fern{\'a}ndez-Hern{\'a}ndez}, {Fernique}, {Fienga}, {Figueras}, {Filippi},
  {Findeisen}, {Fonti}, {Fouesneau}, {Fraile}, {Fraser}, {Fuchs}, {Furnell}, {Gai}, {Galleti}, {Galluccio}, {Garabato}, {Garc{\'\i}a-Sedano}, {Gar{\'e}}, {Garofalo}, {Garralda}, {Gavras}, {Gerssen}, {Geyer}, {Gilmore}, {Girona}, {Giuffrida}, {Gomes}, {Gonz{\'a}lez-Marcos}, {Gonz{\'a}lez-N{\'u}{\~n}ez}, {Gonz{\'a}lez-Vidal}, {Granvik}, {Guerrier}, {Guillout}, {Guiraud}, {G{\'u}rpide}, {Guti{\'e}rrez-S{\'a}nchez}, {Guy}, {Haigron}, {Hatzidimitriou}, {Haywood}, {Heiter}, {Helmi}, {Hobbs}, {Hofmann}, {Holl}, {Holland }, {Hunt}, {Hypki}, {Icardi}, {Irwin}, {Jevardat de Fombelle}, {Jofr{\'e}}, {Jonker}, {Jorissen}, {Julbe}, {Karampelas}, {Kochoska}, {Kohley}, {Kolenberg}, {Kontizas}, {Koposov}, {Kordopatis}, {Koubsky}, {Kowalczyk}, {Krone-Martins}, {Kudryashova}, {Kull}, {Bachchan}, {Lacoste-Seris}, {Lanza}, {Lavigne}, {Le Poncin-Lafitte}, {Lebreton}, {Lebzelter}, {Leccia}, {Leclerc}, {Lecoeur-Taibi}, {Lemaitre}, {Lenhardt}, {Leroux}, {Liao}, {Licata}, {Lindstr{\o}m}, {Lister}, {Livanou}, {Lobel}, {L{\"o}ffler},
  {L{\'o}pez}, {Lopez-Lozano}, {Lorenz}, {Loureiro}, {MacDonald}, {Magalh{\~a}es Fernandes}, {Managau}, {Mann}, {Mantelet}, {Marchal}, {Marchant}, {Marconi}, {Marie}, {Marinoni}, {Marrese}, {Marschalk{\'o}}, {Marshall}, {Mart{\'\i}n-Fleitas}, {Martino}, {Mary}, {Matijevi{\v{c}}}, {Mazeh}, {McMillan}, {Messina}, {Mestre}, {Michalik}, {Millar}, {Miranda}, {Molina}, {Molinaro}, {Molinaro}, {Moln{\'a}r}, {Moniez}, {Montegriffo}, {Monteiro}, {Mor}, {Mora}, {Morbidelli}, {Morel}, {Morgenthaler}, {Morley}, {Morris}, {Mulone}, {Muraveva}, {Musella}, {Narbonne}, {Nelemans}, {Nicastro}, {Noval}, {Ord{\'e}novic}, {Ordieres-Mer{\'e}}, {Osborne}, {Pagani}, {Pagano}, {Pailler}, {Palacin}, {Palaversa}, {Parsons}, {Paulsen}, {Pecoraro}, {Pedrosa}, {Pentik{\"a}inen}, {Pereira}, {Pichon}, {Piersimoni}, {Pineau}, {Plachy}, {Plum}, {Poujoulet}, {Pr{\v{s}}a}, {Pulone}, {Ragaini}, {Rago}, {Rambaux}, {Ramos-Lerate}, {Ranalli}, {Rauw}, {Read}, {Regibo}, {Renk}, {Reyl{\'e}}, {Ribeiro}, {Rimoldini}, {Ripepi}, {Riva}, {Rixon},
  {Roelens}, {Romero-G{\'o}mez}, {Rowell}, {Royer}, {Rudolph}, {Ruiz-Dern}, {Sadowski}, {Sagrist{\`a} Sell{\'e}s}, {Sahlmann}, {Salgado}, {Salguero}, {Sarasso}, {Savietto}, {Schnorhk}, {Schultheis}, {Sciacca}, {Segol}, {Segovia}, {Segransan}, {Serpell}, {Shih}, {Smareglia}, {Smart}, {Smith}, {Solano}, {Solitro}, {Sordo}, {Soria Nieto}, {Souchay}, {Spagna}, {Spoto}, {Stampa}, {Steele}, {Steidelm{\"u}ller}, {Stephenson}, {Stoev}, {Suess}, {S{\"u}veges}, {Surdej}, {Szabados}, {Szegedi-Elek}, {Tapiador}, {Taris}, {Tauran}, {Taylor}, {Teixeira}, {Terrett}, {Tingley}, {Trager}, {Turon}, {Ulla}, {Utrilla}, {Valentini}, {van Elteren}, {Van Hemelryck}, {van Leeuwen}, {Varadi}, {Vecchiato}, {Veljanoski}, {Via}, {Vicente}, {Vogt}, {Voss}, {Votruba}, {Voutsinas}, {Walmsley}, {Weiler}, {Weingrill}, {Werner}, {Wevers}, {Whitehead}, {Wyrzykowski}, {Yoldas}, {{\v{Z}}erjal}, {Zucker}, {Zurbach}, {Zwitter}, {Alecu}, {Allen}, {Allende Prieto}, {Amorim}, {Anglada-Escud{\'e}}, {Arsenijevic}, {Azaz}, {Balm}, {Beck}, {Bernstein},
  {Bigot}, {Bijaoui}, {Blasco}, {Bonfigli}, {Bono}, {Boudreault}, {Bressan}, {Brown}, {Brunet}, {Bunclark}, {Buonanno}, {Butkevich}, {Carret}, {Carrion}, {Chemin}, {Ch{\'e}reau}, {Corcione}, {Darmigny}, {de Boer}, {de Teodoro}, {de Zeeuw}, {Delle Luche}, {Domingues}, {Dubath}, {Fodor}, {Fr{\'e}zouls}, {Fries}, {Fustes}, {Fyfe}, {Gallardo}, {Gallegos}, {Gardiol}, {Gebran}, {Gomboc}, {G{\'o}mez}, {Grux}, {Gueguen}, {Heyrovsky}, {Hoar}, {Iannicola}, {Isasi Parache}, {Janotto}, {Joliet}, {Jonckheere}, {Keil}, {Kim}, {Klagyivik}, {Klar}, {Knude}, {Kochukhov}, {Kolka}, {Kos}, {Kutka}, {Lainey}, {LeBouquin}, {Liu}, {Loreggia}, {Makarov}, {Marseille}, {Martayan}, {Martinez-Rubi}, {Massart}, {Meynadier}, {Mignot}, {Munari}, {Nguyen}, {Nordlander}, {Ocvirk}, {O'Flaherty}, {Olias Sanz}, {Ortiz}, {Osorio}, {Oszkiewicz}, {Ouzounis}, {Palmer}, {Park}, {Pasquato}, {Peltzer}, {Peralta}, {P{\'e}turaud}, {Pieniluoma}, {Pigozzi}, {Poels}, {Prat}, {Prod'homme}, {Raison}, {Rebordao}, {Risquez}, {Rocca-Volmerange}, {Rosen},
  {Ruiz-Fuertes}, {Russo}, {Sembay}, {Serraller Vizcaino}, {Short}, {Siebert}, {Silva}, {Sinachopoulos}, {Slezak}, {Soffel}, {Sosnowska}, {Strai{\v{z}}ys}, {ter Linden}, {Terrell}, {Theil}, {Tiede}, {Troisi}, {Tsalmantza}, {Tur}, {Vaccari}, {Vachier}, {Valles}, {Van Hamme}, {Veltz}, {Virtanen}, {Wallut}, {Wichmann}, {Wilkinson}, {Ziaeepour}, \& {Zschocke}}]{GaiaMission}
{Gaia Collaboration}, {Prusti}, T., {de Bruijne}, J.~H.~J., {et~al.} 2016, \aap, 595, A1, \dodoi{10.1051/0004-6361/201629272}

\bibitem[{{Gaia Collaboration} {et~al.}(2023){Gaia Collaboration}, {Vallenari}, {Brown}, {Prusti}, {de Bruijne}, {Arenou}, {Babusiaux}, {Biermann}, {Creevey}, {Ducourant}, {Evans}, {Eyer}, {Guerra}, {Hutton}, {Jordi}, {Klioner}, {Lammers}, {Lindegren}, {Luri}, {Mignard}, {Panem}, {Pourbaix}, {Randich}, {Sartoretti}, {Soubiran}, {Tanga}, {Walton}, {Bailer-Jones}, {Bastian}, {Drimmel}, {Jansen}, {Katz}, {Lattanzi}, {van Leeuwen}, {Bakker}, {Cacciari}, {Casta{\~n}eda}, {De Angeli}, {Fabricius}, {Fouesneau}, {Fr{\'e}mat}, {Galluccio}, {Guerrier}, {Heiter}, {Masana}, {Messineo}, {Mowlavi}, {Nicolas}, {Nienartowicz}, {Pailler}, {Panuzzo}, {Riclet}, {Roux}, {Seabroke}, {Sordo}, {Th{\'e}venin}, {Gracia-Abril}, {Portell}, {Teyssier}, {Altmann}, {Andrae}, {Audard}, {Bellas-Velidis}, {Benson}, {Berthier}, {Blomme}, {Burgess}, {Busonero}, {Busso}, {C{\'a}novas}, {Carry}, {Cellino}, {Cheek}, {Clementini}, {Damerdji}, {Davidson}, {de Teodoro}, {Nu{\~n}ez Campos}, {Delchambre}, {Dell'Oro}, {Esquej},
  {Fern{\'a}ndez-Hern{\'a}ndez}, {Fraile}, {Garabato}, {Garc{\'\i}a-Lario}, {Gosset}, {Haigron}, {Halbwachs}, {Hambly}, {Harrison}, {Hern{\'a}ndez}, {Hestroffer}, {Hodgkin}, {Holl}, {Jan{\ss}en}, {Jevardat de Fombelle}, {Jordan}, {Krone-Martins}, {Lanzafame}, {L{\"o}ffler}, {Marchal}, {Marrese}, {Moitinho}, {Muinonen}, {Osborne}, {Pancino}, {Pauwels}, {Recio-Blanco}, {Reyl{\'e}}, {Riello}, {Rimoldini}, {Roegiers}, {Rybizki}, {Sarro}, {Siopis}, {Smith}, {Sozzetti}, {Utrilla}, {van Leeuwen}, {Abbas}, {{\'A}brah{\'a}m}, {Abreu Aramburu}, {Aerts}, {Aguado}, {Ajaj}, {Aldea-Montero}, {Altavilla}, {{\'A}lvarez}, {Alves}, {Anders}, {Anderson}, {Anglada Varela}, {Antoja}, {Baines}, {Baker}, {Balaguer-N{\'u}{\~n}ez}, {Balbinot}, {Balog}, {Barache}, {Barbato}, {Barros}, {Barstow}, {Bartolom{\'e}}, {Bassilana}, {Bauchet}, {Becciani}, {Bellazzini}, {Berihuete}, {Bernet}, {Bertone}, {Bianchi}, {Binnenfeld}, {Blanco-Cuaresma}, {Blazere}, {Boch}, {Bombrun}, {Bossini}, {Bouquillon}, {Bragaglia}, {Bramante}, {Breedt},
  {Bressan}, {Brouillet}, {Brugaletta}, {Bucciarelli}, {Burlacu}, {Butkevich}, {Buzzi}, {Caffau}, {Cancelliere}, {Cantat-Gaudin}, {Carballo}, {Carlucci}, {Carnerero}, {Carrasco}, {Casamiquela}, {Castellani}, {Castro-Ginard}, {Chaoul}, {Charlot}, {Chemin}, {Chiaramida}, {Chiavassa}, {Chornay}, {Comoretto}, {Contursi}, {Cooper}, {Cornez}, {Cowell}, {Crifo}, {Cropper}, {Crosta}, {Crowley}, {Dafonte}, {Dapergolas}, {David}, {David}, {de Laverny}, {De Luise}, {De March}, {De Ridder}, {de Souza}, {de Torres}, {del Peloso}, {del Pozo}, {Delbo}, {Delgado}, {Delisle}, {Demouchy}, {Dharmawardena}, {Di Matteo}, {Diakite}, {Diener}, {Distefano}, {Dolding}, {Edvardsson}, {Enke}, {Fabre}, {Fabrizio}, {Faigler}, {Fedorets}, {Fernique}, {Fienga}, {Figueras}, {Fournier}, {Fouron}, {Fragkoudi}, {Gai}, {Garcia-Gutierrez}, {Garcia-Reinaldos}, {Garc{\'\i}a-Torres}, {Garofalo}, {Gavel}, {Gavras}, {Gerlach}, {Geyer}, {Giacobbe}, {Gilmore}, {Girona}, {Giuffrida}, {Gomel}, {Gomez}, {Gonz{\'a}lez-N{\'u}{\~n}ez},
  {Gonz{\'a}lez-Santamar{\'\i}a}, {Gonz{\'a}lez-Vidal}, {Granvik}, {Guillout}, {Guiraud}, {Guti{\'e}rrez-S{\'a}nchez}, {Guy}, {Hatzidimitriou}, {Hauser}, {Haywood}, {Helmer}, {Helmi}, {Sarmiento}, {Hidalgo}, {Hilger}, {H{\l}adczuk}, {Hobbs}, {Holland}, {Huckle}, {Jardine}, {Jasniewicz}, {Jean-Antoine Piccolo}, {Jim{\'e}nez-Arranz}, {Jorissen}, {Juaristi Campillo}, {Julbe}, {Karbevska}, {Kervella}, {Khanna}, {Kontizas}, {Kordopatis}, {Korn}, {K{\'o}sp{\'a}l}, {Kostrzewa-Rutkowska}, {Kruszy{\'n}ska}, {Kun}, {Laizeau}, {Lambert}, {Lanza}, {Lasne}, {Le Campion}, {Lebreton}, {Lebzelter}, {Leccia}, {Leclerc}, {Lecoeur-Taibi}, {Liao}, {Licata}, {Lindstr{\o}m}, {Lister}, {Livanou}, {Lobel}, {Lorca}, {Loup}, {Madrero Pardo}, {Magdaleno Romeo}, {Managau}, {Mann}, {Manteiga}, {Marchant}, {Marconi}, {Marcos}, {Marcos Santos}, {Mar{\'\i}n Pina}, {Marinoni}, {Marocco}, {Marshall}, {Martin Polo}, {Mart{\'\i}n-Fleitas}, {Marton}, {Mary}, {Masip}, {Massari}, {Mastrobuono-Battisti}, {Mazeh}, {McMillan}, {Messina}, {Michalik},
  {Millar}, {Mints}, {Molina}, {Molinaro}, {Moln{\'a}r}, {Monari}, {Mongui{\'o}}, {Montegriffo}, {Montero}, {Mor}, {Mora}, {Morbidelli}, {Morel}, {Morris}, {Muraveva}, {Murphy}, {Musella}, {Nagy}, {Noval}, {Oca{\~n}a}, {Ogden}, {Ordenovic}, {Osinde}, {Pagani}, {Pagano}, {Palaversa}, {Palicio}, {Pallas-Quintela}, {Panahi}, {Payne-Wardenaar}, {Pe{\~n}alosa Esteller}, {Penttil{\"a}}, {Pichon}, {Piersimoni}, {Pineau}, {Plachy}, {Plum}, {Poggio}, {Pr{\v{s}}a}, {Pulone}, {Racero}, {Ragaini}, {Rainer}, {Raiteri}, {Rambaux}, {Ramos}, {Ramos-Lerate}, {Re Fiorentin}, {Regibo}, {Richards}, {Rios Diaz}, {Ripepi}, {Riva}, {Rix}, {Rixon}, {Robichon}, {Robin}, {Robin}, {Roelens}, {Rogues}, {Rohrbasser}, {Romero-G{\'o}mez}, {Rowell}, {Royer}, {Ruz Mieres}, {Rybicki}, {Sadowski}, {S{\'a}ez N{\'u}{\~n}ez}, {Sagrist{\`a} Sell{\'e}s}, {Sahlmann}, {Salguero}, {Samaras}, {Sanchez Gimenez}, {Sanna}, {Santove{\~n}a}, {Sarasso}, {Schultheis}, {Sciacca}, {Segol}, {Segovia}, {S{\'e}gransan}, {Semeux}, {Shahaf}, {Siddiqui}, {Siebert},
  {Siltala}, {Silvelo}, {Slezak}, {Slezak}, {Smart}, {Snaith}, {Solano}, {Solitro}, {Souami}, {Souchay}, {Spagna}, {Spina}, {Spoto}, {Steele}, {Steidelm{\"u}ller}, {Stephenson}, {S{\"u}veges}, {Surdej}, {Szabados}, {Szegedi-Elek}, {Taris}, {Taylor}, {Teixeira}, {Tolomei}, {Tonello}, {Torra}, {Torra}, {Torralba Elipe}, {Trabucchi}, {Tsounis}, {Turon}, {Ulla}, {Unger}, {Vaillant}, {van Dillen}, {van Reeven}, {Vanel}, {Vecchiato}, {Viala}, {Vicente}, {Voutsinas}, {Weiler}, {Wevers}, {Wyrzykowski}, {Yoldas}, {Yvard}, {Zhao}, {Zorec}, {Zucker}, \& {Zwitter}}]{GAIADR32023}
{Gaia Collaboration}, {Vallenari}, A., {Brown}, A.~G.~A., {et~al.} 2023, \aap, 674, A1, \dodoi{10.1051/0004-6361/202243940}

\bibitem[{{Gallart} {et~al.}(2019){Gallart}, {Bernard}, {Brook}, {Ruiz-Lara}, {Cassisi}, {Hill}, \& {Monelli}}]{Gallart2019}
{Gallart}, C., {Bernard}, E.~J., {Brook}, C.~B., {et~al.} 2019, Nature Astronomy, 3, 932, \dodoi{10.1038/s41550-019-0829-5}

\bibitem[{{Gallart} {et~al.}(2024){Gallart}, {Surot}, {Cassisi}, {Fern{\'a}ndez-Alvar}, {Mirabal}, {Rivero}, {Ruiz-Lara}, {Santos-Torres}, {Aznar-Menargues}, {Battaglia}, {Queiroz}, {Monelli}, {Vasiliev}, {Chiappini}, {Helmi}, {Hill}, {Massari}, \& {Thomas}}]{Gallart2024}
{Gallart}, C., {Surot}, F., {Cassisi}, S., {et~al.} 2024, arXiv e-prints, arXiv:2402.09399, \dodoi{10.48550/arXiv.2402.09399}

\bibitem[{{Garc{\'\i}a P{\'e}rez} {et~al.}(2016){Garc{\'\i}a P{\'e}rez}, {Allende Prieto}, {Holtzman}, {Shetrone}, {M{\'e}sz{\'a}ros}, {Bizyaev}, {Carrera}, {Cunha}, {Garc{\'\i}a-Hern{\'a}ndez}, {Johnson}, {Majewski}, {Nidever}, {Schiavon}, {Shane}, {Smith}, {Sobeck}, {Troup}, {Zamora}, {Weinberg}, {Bovy}, {Eisenstein}, {Feuillet}, {Frinchaboy}, {Hayden}, {Hearty}, {Nguyen}, {O'Connell}, {Pinsonneault}, {Wilson}, \& {Zasowski}}]{GarciaPerez2016}
{Garc{\'\i}a P{\'e}rez}, A.~E., {Allende Prieto}, C., {Holtzman}, J.~A., {et~al.} 2016, \aj, 151, 144, \dodoi{10.3847/0004-6256/151/6/144}

\bibitem[{{Garver} {et~al.}(2023){Garver}, {Nidever}, {Debattista}, {Beraldo e Silva}, \& {Khachaturyants}}]{Garver2023}
{Garver}, B.~R., {Nidever}, D.~L., {Debattista}, V.~P., {Beraldo e Silva}, L., \& {Khachaturyants}, T. 2023, \apj, 953, 128, \dodoi{10.3847/1538-4357/acdfc6}

\bibitem[{{Gilmore} \& {Reid}(1983)}]{Gilmore1983}
{Gilmore}, G., \& {Reid}, N. 1983, \mnras, 202, 1025.
\newblock \url{http://adsabs.harvard.edu/abs/1983MNRAS.202.1025G}

\bibitem[{{Giribaldi} \& {Smiljanic}(2023)}]{Giribaldi2023}
{Giribaldi}, R.~E., \& {Smiljanic}, R. 2023, \aap, 673, A18, \dodoi{10.1051/0004-6361/202245404}

\bibitem[{{Grand} {et~al.}(2020){Grand}, {Kawata}, {Belokurov}, {Deason}, {Fattahi}, {Fragkoudi}, {G{\'o}mez}, {Marinacci}, \& {Pakmor}}]{Grand2020}
{Grand}, R. J.~J., {Kawata}, D., {Belokurov}, V., {et~al.} 2020, \mnras, 497, 1603, \dodoi{10.1093/mnras/staa2057}

\bibitem[{{Grisoni} {et~al.}(2021){Grisoni}, {Matteucci}, \& {Romano}}]{Grisoni2021}
{Grisoni}, V., {Matteucci}, F., \& {Romano}, D. 2021, \mnras, 508, 719, \dodoi{10.1093/mnras/stab2579}

\bibitem[{{Grisoni} {et~al.}(2017){Grisoni}, {Spitoni}, {Matteucci}, {Recio-Blanco}, {de Laverny}, {Hayden}, {Mikolaitis}, \& {Worley}}]{Grisoni2017}
{Grisoni}, V., {Spitoni}, E., {Matteucci}, F., {et~al.} 2017, \mnras, 472, 3637, \dodoi{10.1093/mnras/stx2201}

\bibitem[{{Grisoni} {et~al.}(2024){Grisoni}, {Chiappini}, {Miglio}, {Brogaard}, {Casali}, {Willett}, {Montalb{\'a}n}, {Stokholm}, {Thomsen}, {Tailo}, {Matteuzzi}, {Valentini}, {Elsworth}, \& {Mosser}}]{Grisoni2024}
{Grisoni}, V., {Chiappini}, C., {Miglio}, A., {et~al.} 2024, \aap, 683, A111, \dodoi{10.1051/0004-6361/202347440}

\bibitem[{{Guiglion} {et~al.}(2024){Guiglion}, {Nepal}, {Chiappini}, {Khoperskov}, {Traven}, {Queiroz}, {Steinmetz}, {Valentini}, {Fournier}, {Vallenari}, {Youakim}, {Bergemann}, {M{\'e}sz{\'a}ros}, {Lucatello}, {Sordo}, {Fabbro}, {Minchev}, {Tautvai{\v{s}}ien{\.{e}}}, {Mikolaitis}, \& {Montalb{\'a}n}}]{Guiglion2024}
{Guiglion}, G., {Nepal}, S., {Chiappini}, C., {et~al.} 2024, \aap, 682, A9, \dodoi{10.1051/0004-6361/202347122}

\bibitem[{{Gunn} {et~al.}(2006){Gunn}, {Siegmund}, {Mannery}, {Owen}, {Hull}, {Leger}, {Carey}, {Knapp}, {York}, {Boroski}, {Kent}, {Lupton}, {Rockosi}, {Evans}, {Waddell}, {Anderson}, {Annis}, {Barentine}, {Bartoszek}, {Bastian}, {Bracker}, {Brewington}, {Briegel}, {Brinkmann}, {Brown}, {Carr}, {Czarapata}, {Drennan}, {Dombeck}, {Federwitz}, {Gillespie}, {Gonzales}, {Hansen}, {Harvanek}, {Hayes}, {Jordan}, {Kinney}, {Klaene}, {Kleinman}, {Kron}, {Kresinski}, {Lee}, {Limmongkol}, {Lindenmeyer}, {Long}, {Loomis}, {McGehee}, {Mantsch}, {Neilsen}, {Neswold}, {Newman}, {Nitta}, {Peoples}, {Pier}, {Prieto}, {Prosapio}, {Rivetta}, {Schneider}, {Snedden}, \& {Wang}}]{Gunn2006}
{Gunn}, J.~E., {Siegmund}, W.~A., {Mannery}, E.~J., {et~al.} 2006, \aj, 131, 2332, \dodoi{10.1086/500975}

\bibitem[{{Hasselquist} {et~al.}(2021){Hasselquist}, {Hayes}, {Lian}, {Weinberg}, {Zasowski}, {Horta}, {Beaton}, {Feuillet}, {Garro}, {Gallart}, {Smith}, {Holtzman}, {Minniti}, {Lacerna}, {Shetrone}, {J{\"o}nsson}, {Cioni}, {Fillingham}, {Cunha}, {O'Connell}, {Fern{\'a}ndez-Trincado}, {Mu{\~n}oz}, {Schiavon}, {Almeida}, {Anguiano}, {Beers}, {Bizyaev}, {Brownstein}, {Cohen}, {Frinchaboy}, {Garc{\'\i}a-Hern{\'a}ndez}, {Geisler}, {Lane}, {Majewski}, {Nidever}, {Nitschelm}, {Povick}, {Price-Whelan}, {Roman-Lopes}, {Rosado}, {Sobeck}, {Stringfellow}, {Valenzuela}, {Villanova}, \& {Vincenzo}}]{Hasselquist2021}
{Hasselquist}, S., {Hayes}, C.~R., {Lian}, J., {et~al.} 2021, \apj, 923, 172, \dodoi{10.3847/1538-4357/ac25f9}

\bibitem[{{Hayden} {et~al.}(2015){Hayden}, {Bovy}, {Holtzman}, {Nidever}, {Bird}, {Weinberg}, {Andrews}, {Majewski}, {Allende Prieto}, {Anders}, {Beers}, {Bizyaev}, {Chiappini}, {Cunha}, {Frinchaboy}, {Garc{\'\i}a-Her{\'n}andez}, {Garc{\'\i}a P{\'e}rez}, {Girardi}, {Harding}, {Hearty}, {Johnson}, {M{\'e}sz{\'a}ros}, {Minchev}, {O'Connell}, {Pan}, {Robin}, {Schiavon}, {Schneider}, {Schultheis}, {Shetrone}, {Skrutskie}, {Steinmetz}, {Smith}, {Wilson}, {Zamora}, \& {Zasowski}}]{Hayden2015}
{Hayden}, M.~R., {Bovy}, J., {Holtzman}, J.~A., {et~al.} 2015, \apj, 808, 132, \dodoi{10.1088/0004-637X/808/2/132}

\bibitem[{{Haywood} {et~al.}(2013){Haywood}, {Di Matteo}, {Lehnert}, {Katz}, \& {G{\'o}mez}}]{Haywood2013}
{Haywood}, M., {Di Matteo}, P., {Lehnert}, M.~D., {Katz}, D., \& {G{\'o}mez}, A. 2013, \aap, 560, A109, \dodoi{10.1051/0004-6361/201321397}

\bibitem[{{Haywood} {et~al.}(2018){Haywood}, {Di Matteo}, {Lehnert}, {Snaith}, {Khoperskov}, \& {G{\'o}mez}}]{Haywood2018}
{Haywood}, M., {Di Matteo}, P., {Lehnert}, M.~D., {et~al.} 2018, \apj, 863, 113, \dodoi{10.3847/1538-4357/aad235}

\bibitem[{{Helmi} {et~al.}(2018){Helmi}, {Babusiaux}, {Koppelman}, {Massari}, {Veljanoski}, \& {Brown}}]{helmi2018}
{Helmi}, A., {Babusiaux}, C., {Koppelman}, H.~H., {et~al.} 2018, \nat, 563, 85, \dodoi{10.1038/s41586-018-0625-x}

\bibitem[{{Howell}(2011)}]{Howell2011}
{Howell}, D.~A. 2011, Nature Communications, 2, 350, \dodoi{10.1038/ncomms1344}

\bibitem[{{Imig} {et~al.}(2023){Imig}, {Price}, {Holtzman}, {Stone-Martinez}, {Majewski}, {Weinberg}, {Johnson}, {Allende Prieto}, {Beaton}, {Beers}, {Bizyaev}, {Blanton}, {Brownstein}, {Cunha}, {Fern{\'a}ndez-Trincado}, {Feuillet}, {Hasselquist}, {Hayes}, {J{\"o}nsson}, {Lane}, {Lian}, {M{\'e}sz{\'a}ros}, {Nidever}, {Robin}, {Shetrone}, {Smith}, \& {Wilson}}]{Imig2023}
{Imig}, J., {Price}, C., {Holtzman}, J.~A., {et~al.} 2023, \apj, 954, 124, \dodoi{10.3847/1538-4357/ace9b8}

\bibitem[{{Izzard} {et~al.}(2018){Izzard}, {Preece}, {Jofre}, {Halabi}, {Masseron}, \& {Tout}}]{Izzard2018}
{Izzard}, R.~G., {Preece}, H., {Jofre}, P., {et~al.} 2018, \mnras, 473, 2984, \dodoi{10.1093/mnras/stx2355}

\bibitem[{{Jofr{\'e}} {et~al.}(2016){Jofr{\'e}}, {Jorissen}, {Van Eck}, {Izzard}, {Masseron}, {Hawkins}, {Gilmore}, {Paladini}, {Escorza}, {Blanco-Cuaresma}, \& {Manick}}]{Jofre2016}
{Jofr{\'e}}, P., {Jorissen}, A., {Van Eck}, S., {et~al.} 2016, \aap, 595, A60, \dodoi{10.1051/0004-6361/201629356}

\bibitem[{{Jofr{\'e}} {et~al.}(2023){Jofr{\'e}}, {Jorissen}, {Aguilera-G{\'o}mez}, {Van Eck}, {Tayar}, {Pinsonneault}, {Zinn}, {Goriely}, \& {Van Winckel}}]{Jofre2023}
{Jofr{\'e}}, P., {Jorissen}, A., {Aguilera-G{\'o}mez}, C., {et~al.} 2023, \aap, 671, A21, \dodoi{10.1051/0004-6361/202244524}

\bibitem[{{Johnson} {et~al.}(2018){Johnson}, {Harrison}, {Swinbank}, {Tiley}, {Stott}, {Bower}, {Smail}, {Bunker}, {Sobral}, {Turner}, {Best}, {Bureau}, {Cirasuolo}, {Jarvis}, {Magdis}, {Sharples}, {Bland-Hawthorn}, {Catinella}, {Cortese}, {Croom}, {Federrath}, {Glazebrook}, {Sweet}, {Bryant}, {Goodwin}, {Konstantopoulos}, {Lawrence}, {Medling}, {Owers}, \& {Richards}}]{Johnson2018}
{Johnson}, H.~L., {Harrison}, C.~M., {Swinbank}, A.~M., {et~al.} 2018, \mnras, 474, 5076, \dodoi{10.1093/mnras/stx3016}

\bibitem[{{J{\"o}nsson} {et~al.}(2020){J{\"o}nsson}, {Holtzman}, {Allende Prieto}, {Cunha}, {Garc{\'\i}a-Hern{\'a}ndez}, {Hasselquist}, {Masseron}, {Osorio}, {Shetrone}, {Smith}, {Stringfellow}, {Bizyaev}, {Edvardsson}, {Majewski}, {M{\'e}sz{\'a}ros}, {Souto}, {Zamora}, {Beaton}, {Bovy}, {Donor}, {Pinsonneault}, {Poovelil}, \& {Sobeck}}]{jonsson2020}
{J{\"o}nsson}, H., {Holtzman}, J.~A., {Allende Prieto}, C., {et~al.} 2020, \aj, 160, 120, \dodoi{10.3847/1538-3881/aba592}

\bibitem[{{Juri{\'c}} {et~al.}(2008){Juri{\'c}}, {Ivezi{\'c}}, {Brooks}, {Lupton}, {Schlegel}, {Finkbeiner}, {Padmanabhan}, {Bond}, {Sesar}, {Rockosi}, {Knapp}, {Gunn}, {Sumi}, {Schneider}, {Barentine}, {Brewington}, {Brinkmann}, {Fukugita}, {Harvanek}, {Kleinman}, {Krzesinski}, {Long}, {Neilsen}, {Nitta}, {Snedden}, \& {York}}]{Juric2008}
{Juri{\'c}}, M., {Ivezi{\'c}}, {\v{Z}}., {Brooks}, A., {et~al.} 2008, \apj, 673, 864, \dodoi{10.1086/523619}

\bibitem[{{Kordopatis} {et~al.}(2013){Kordopatis}, {Gilmore}, {Wyse}, {Steinmetz}, {Siebert}, {Bienaym{\'e}}, {McMillan}, {Minchev}, {Zwitter}, {Gibson}, {Seabroke}, {Grebel}, {Bland-Hawthorn}, {Boeche}, {Freeman}, {Munari}, {Navarro}, {Parker}, {Reid}, \& {Siviero}}]{Kordopatis2013}
{Kordopatis}, G., {Gilmore}, G., {Wyse}, R.~F.~G., {et~al.} 2013, \mnras, 436, 3231, \dodoi{10.1093/mnras/stt1804}

\bibitem[{{Lagarde} {et~al.}(2021){Lagarde}, {Reyl{\'e}}, {Chiappini}, {Mor}, {Anders}, {Figueras}, {Miglio}, {Romero-G{\'o}mez}, {Antoja}, {Cabral}, {Salomon}, {Robin}, {Bienaym{\'e}}, {Soubiran}, {Cornu}, \& {Montillaud}}]{Lagarde2021}
{Lagarde}, N., {Reyl{\'e}}, C., {Chiappini}, C., {et~al.} 2021, \aap, 654, A13, \dodoi{10.1051/0004-6361/202039982}

\bibitem[{{Laporte} {et~al.}(2020){Laporte}, {Famaey}, {Monari}, {Hill}, {Wegg}, \& {Gerhard}}]{Laporte2020}
{Laporte}, C. F.~P., {Famaey}, B., {Monari}, G., {et~al.} 2020, \aap, 643, L3, \dodoi{10.1051/0004-6361/202038740}

\bibitem[{{Lee} {et~al.}(2011){Lee}, {Beers}, {An}, {Ivezi{\'c}}, {Just}, {Rockosi}, {Morrison}, {Johnson}, {Sch{\"o}nrich}, {Bird}, {Yanny}, {Harding}, \& {Rocha-Pinto}}]{Lee2011}
{Lee}, Y.~S., {Beers}, T.~C., {An}, D., {et~al.} 2011, \apj, 738, 187, \dodoi{10.1088/0004-637X/738/2/187}

\bibitem[{{Li} \& {Zhao}(2017)}]{LiZhao2017}
{Li}, C., \& {Zhao}, G. 2017, \apj, 850, 25, \dodoi{10.3847/1538-4357/aa93f4}

\bibitem[{{Limberg} {et~al.}(2022){Limberg}, {Souza}, {P{\'e}rez-Villegas}, {Rossi}, {Perottoni}, \& {Santucci}}]{limberg2022}
{Limberg}, G., {Souza}, S.~O., {P{\'e}rez-Villegas}, A., {et~al.} 2022, \apj, 935, 109, \dodoi{10.3847/1538-4357/ac8159}

\bibitem[{{Lindegren} {et~al.}(2021){Lindegren}, {Klioner}, {Hern{\'a}ndez}, {Bombrun}, {Ramos-Lerate}, {Steidelm{\"u}ller}, {Bastian}, {Biermann}, {de Torres}, {Gerlach}, {Geyer}, {Hilger}, {Hobbs}, {Lammers}, {McMillan}, {Stephenson}, {Casta{\~n}eda}, {Davidson}, {Fabricius}, {Gracia-Abril}, {Portell}, {Rowell}, {Teyssier}, {Torra}, {Bartolom{\'e}}, {Clotet}, {Garralda}, {Gonz{\'a}lez-Vidal}, {Torra}, {Abbas}, {Altmann}, {Anglada Varela}, {Balaguer-N{\'u}{\~n}ez}, {Balog}, {Barache}, {Becciani}, {Bernet}, {Bertone}, {Bianchi}, {Bouquillon}, {Brown}, {Bucciarelli}, {Busonero}, {Butkevich}, {Buzzi}, {Cancelliere}, {Carlucci}, {Charlot}, {Cioni}, {Crosta}, {Crowley}, {del Peloso}, {del Pozo}, {Drimmel}, {Esquej}, {Fienga}, {Fraile}, {Gai}, {Garcia-Reinaldos}, {Guerra}, {Hambly}, {Hauser}, {Jan{\ss}en}, {Jordan}, {Kostrzewa-Rutkowska}, {Lattanzi}, {Liao}, {Licata}, {Lister}, {L{\"o}ffler}, {Marchant}, {Masip}, {Mignard}, {Mints}, {Molina}, {Mora}, {Morbidelli}, {Murphy}, {Pagani}, {Panuzzo}, {Pe{\~n}alosa
  Esteller}, {Poggio}, {Re Fiorentin}, {Riva}, {Sagrist{\`a} Sell{\'e}s}, {Sanchez Gimenez}, {Sarasso}, {Sciacca}, {Siddiqui}, {Smart}, {Souami}, {Spagna}, {Steele}, {Taris}, {Utrilla}, {van Reeven}, \& {Vecchiato}}]{Lindegren2020a}
{Lindegren}, L., {Klioner}, S.~A., {Hern{\'a}ndez}, J., {et~al.} 2021, \aap, 649, A2, \dodoi{10.1051/0004-6361/202039709}

\bibitem[{{Majewski} {et~al.}(2017){Majewski}, {Schiavon}, {Frinchaboy}, {Allende Prieto}, {Barkhouser}, {Bizyaev}, {Blank}, {Brunner}, {Burton}, {Carrera}, {Chojnowski}, {Cunha}, {Epstein}, {Fitzgerald}, {Garc{\'\i}a P{\'e}rez}, {Hearty}, {Henderson}, {Holtzman}, {Johnson}, {Lam}, {Lawler}, {Maseman}, {M{\'e}sz{\'a}ros}, {Nelson}, {Nguyen}, {Nidever}, {Pinsonneault}, {Shetrone}, {Smee}, {Smith}, {Stolberg}, {Skrutskie}, {Walker}, {Wilson}, {Zasowski}, {Anders}, {Basu}, {Beland}, {Blanton}, {Bovy}, {Brownstein}, {Carlberg}, {Chaplin}, {Chiappini}, {Eisenstein}, {Elsworth}, {Feuillet}, {Fleming}, {Galbraith-Frew}, {Garc{\'\i}a}, {Garc{\'\i}a-Hern{\'a}ndez}, {Gillespie}, {Girardi}, {Gunn}, {Hasselquist}, {Hayden}, {Hekker}, {Ivans}, {Kinemuchi}, {Klaene}, {Mahadevan}, {Mathur}, {Mosser}, {Muna}, {Munn}, {Nichol}, {O'Connell}, {Parejko}, {Robin}, {Rocha-Pinto}, {Schultheis}, {Serenelli}, {Shane}, {Silva Aguirre}, {Sobeck}, {Thompson}, {Troup}, {Weinberg}, \& {Zamora}}]{APOGEE2017}
{Majewski}, S.~R., {Schiavon}, R.~P., {Frinchaboy}, P.~M., {et~al.} 2017, \aj, 154, 94, \dodoi{10.3847/1538-3881/aa784d}

\bibitem[{{Martig} {et~al.}(2015){Martig}, {Rix}, {Silva Aguirre}, {Hekker}, {Mosser}, {Elsworth}, {Bovy}, {Stello}, {Anders}, {Garc{\'\i}a}, {Tayar}, {Rodrigues}, {Basu}, {Carrera}, {Ceillier}, {Chaplin}, {Chiappini}, {Frinchaboy}, {Garc{\'\i}a-Hern{\'a}ndez}, {Hearty}, {Holtzman}, {Johnson}, {Majewski}, {Mathur}, {M{\'e}sz{\'a}ros}, {Miglio}, {Nidever}, {Pan}, {Pinsonneault}, {Schiavon}, {Schneider}, {Serenelli}, {Shetrone}, \& {Zamora}}]{Martig2015}
{Martig}, M., {Rix}, H.-W., {Silva Aguirre}, V., {et~al.} 2015, \mnras, 451, 2230, \dodoi{10.1093/mnras/stv1071}

\bibitem[{{Matteucci}(2021)}]{Matteucci2021}
{Matteucci}, F. 2021, \aapr, 29, 5, \dodoi{10.1007/s00159-021-00133-8}

\bibitem[{{Matteucci} \& {Greggio}(1986)}]{Matteucci1986}
{Matteucci}, F., \& {Greggio}, L. 1986, \aap, 154, 279

\bibitem[{{McMillan}(2017)}]{mcmillan2017}
{McMillan}, P.~J. 2017, \mnras, 465, 76, \dodoi{10.1093/mnras/stw2759}

\bibitem[{{Miglio} {et~al.}(2021){Miglio}, {Chiappini}, {Mackereth}, {Davies}, {Brogaard}, {Casagrande}, {Chaplin}, {Girardi}, {Kawata}, {Khan}, {Izzard}, {Montalb{\'a}n}, {Mosser}, {Vincenzo}, {Bossini}, {Noels}, {Rodrigues}, {Valentini}, \& {Mandel}}]{Miglio2021}
{Miglio}, A., {Chiappini}, C., {Mackereth}, J.~T., {et~al.} 2021, \aap, 645, A85, \dodoi{10.1051/0004-6361/202038307}

\bibitem[{{Minchev} {et~al.}(2013){Minchev}, {Chiappini}, \& {Martig}}]{Minchev2013}
{Minchev}, I., {Chiappini}, C., \& {Martig}, M. 2013, \aap, 558, A9, \dodoi{10.1051/0004-6361/201220189}

\bibitem[{{Montalb{\'a}n} {et~al.}(2021){Montalb{\'a}n}, {Mackereth}, {Miglio}, {Vincenzo}, {Chiappini}, {Buldgen}, {Mosser}, {Noels}, {Scuflaire}, {Vrard}, {Willett}, {Davies}, {Hall}, {Nielsen}, {Khan}, {Rendle}, {van Rossem}, {Ferguson}, \& {Chaplin}}]{Montalban2021}
{Montalb{\'a}n}, J., {Mackereth}, J.~T., {Miglio}, A., {et~al.} 2021, Nature Astronomy, 5, 640, \dodoi{10.1038/s41550-021-01347-7}

\bibitem[{{Naidu} {et~al.}(2021){Naidu}, {Conroy}, {Bonaca}, {Zaritsky}, {Weinberger}, {Ting}, {Caldwell}, {Tacchella}, {Han}, {Speagle}, \& {Cargile}}]{Naidu2021}
{Naidu}, R.~P., {Conroy}, C., {Bonaca}, A., {et~al.} 2021, \apj, 923, 92, \dodoi{10.3847/1538-4357/ac2d2d}

\bibitem[{{Nepal} {et~al.}(2024{\natexlab{a}}){Nepal}, {Chiappini}, {Queiroz}, {Guiglion}, {Montalb{\'a}n}, {Steinmetz}, {Miglio}, \& {Khalatyan}}]{Nepal2024}
{Nepal}, S., {Chiappini}, C., {Queiroz}, A.~B., {et~al.} 2024{\natexlab{a}}, \aap, 688, A167, \dodoi{10.1051/0004-6361/202449445}

\bibitem[{{Nepal} {et~al.}(2024{\natexlab{b}}){Nepal}, {Chiappini}, {Guiglion}, {Steinmetz}, {P{\'e}rez-Villegas}, {Queiroz}, {Miglio}, {Dohme}, \& {Khalatyan}}]{Nepal2024a}
{Nepal}, S., {Chiappini}, C., {Guiglion}, G., {et~al.} 2024{\natexlab{b}}, \aap, 681, L8, \dodoi{10.1051/0004-6361/202348365}

\bibitem[{{Prudil} {et~al.}(2020){Prudil}, {D{\'e}k{\'a}ny}, {Grebel}, \& {Kunder}}]{Prudil2020}
{Prudil}, Z., {D{\'e}k{\'a}ny}, I., {Grebel}, E.~K., \& {Kunder}, A. 2020, \mnras, 492, 3408, \dodoi{10.1093/mnras/staa046}

\bibitem[{{Queiroz} {et~al.}(2018){Queiroz}, {Anders}, {Santiago}, {Chiappini}, {Steinmetz}, {Dal Ponte}, {Stassun}, {da Costa}, {Maia}, {Crestani}, {Beers}, {Fern{\'a}ndez-Trincado}, {Garc{\'\i}a-Hern{\'a}ndez}, {Roman-Lopes}, \& {Zamora}}]{Queiroz2018}
{Queiroz}, A.~B.~A., {Anders}, F., {Santiago}, B.~X., {et~al.} 2018, \mnras, 476, 2556, \dodoi{10.1093/mnras/sty330}

\bibitem[{{Queiroz} {et~al.}(2020){Queiroz}, {Anders}, {Chiappini}, {Khalatyan}, {Santiago}, {Steinmetz}, {Valentini}, {Miglio}, {Bossini}, {Barbuy}, {Minchev}, {Minniti}, {Garc{\'\i}a Hern{\'a}ndez}, {Schultheis}, {Beaton}, {Beers}, {Bizyaev}, {Brownstein}, {Cunha}, {Fern{\'a}ndez-Trincado}, {Frinchaboy}, {Lane}, {Majewski}, {Nataf}, {Nitschelm}, {Pan}, {Roman-Lopes}, {Sobeck}, {Stringfellow}, \& {Zamora}}]{Queiroz2020}
{Queiroz}, A.~B.~A., {Anders}, F., {Chiappini}, C., {et~al.} 2020, \aap, 638, A76, \dodoi{10.1051/0004-6361/201937364}

\bibitem[{{Queiroz} {et~al.}(2023){Queiroz}, {Anders}, {Chiappini}, {Khalatyan}, {Santiago}, {Nepal}, {Steinmetz}, {Gallart}, {Valentini}, {Dal Ponte}, {Barbuy}, {P{\'e}rez-Villegas}, {Masseron}, {Fern{\'a}ndez-Trincado}, {Khoperskov}, {Minchev}, {Fern{\'a}ndez-Alvar}, {Lane}, \& {Nitschelm}}]{Queiroz2023}
---. 2023, \aap, 673, A155, \dodoi{10.1051/0004-6361/202245399}

\bibitem[{{Renaud} {et~al.}(2021){Renaud}, {Agertz}, {Andersson}, {Read}, {Ryde}, {Bensby}, {Rey}, \& {Feuillet}}]{Renaud2021}
{Renaud}, F., {Agertz}, O., {Andersson}, E.~P., {et~al.} 2021, \mnras, 503, 5868, \dodoi{10.1093/mnras/stab543}

\bibitem[{{Rockosi} {et~al.}(2022){Rockosi}, {Lee}, {Morrison}, {Yanny}, {Johnson}, {Lucatello}, {Sobeck}, {Beers}, {Allende Prieto}, {An}, {Bizyaev}, {Blanton}, {Casagrande}, {Eisenstein}, {Gould}, {Gunn}, {Harding}, {Ivans}, {Jacobson}, {Janesh}, {Knapp}, {Kollmeier}, {L{\'e}pine}, {L{\'o}pez-Corredoira}, {Ma}, {Newberg}, {Pan}, {Prchlik}, {Sayers}, {Schlesinger}, {Simmerer}, \& {Weinberg}}]{SEGUE2}
{Rockosi}, C.~M., {Lee}, Y.~S., {Morrison}, H.~L., {et~al.} 2022, \apjs, 259, 60, \dodoi{10.3847/1538-4365/ac5323}

\bibitem[{{Sales} {et~al.}(2009){Sales}, {Helmi}, {Abadi}, {Brook}, {G{\'o}mez}, {Ro{\v{s}}kar}, {Debattista}, {House}, {Steinmetz}, \& {Villalobos}}]{Sales2009}
{Sales}, L.~V., {Helmi}, A., {Abadi}, M.~G., {et~al.} 2009, \mnras, 400, L61, \dodoi{10.1111/j.1745-3933.2009.00763.x}

\bibitem[{{Sanders} \& {Das}(2018)}]{Sanders&Das2018}
{Sanders}, J.~L., \& {Das}, P. 2018, \mnras, 481, 4093, \dodoi{10.1093/mnras/sty2490}

\bibitem[{{Santiago} {et~al.}(2016){Santiago}, {Brauer}, {Anders}, {Chiappini}, {Queiroz}, {Girardi}, {Rocha-Pinto}, {Balbinot}, {da Costa}, {Maia}, {Schultheis}, {Steinmetz}, {Miglio}, {Montalb{\'a}n}, {Schneider}, {Beers}, {Frinchaboy}, {Lee}, \& {Zasowski}}]{Santiago2016starhorse}
{Santiago}, B.~X., {Brauer}, D.~E., {Anders}, F., {et~al.} 2016, \aap, 585, A42, \dodoi{10.1051/0004-6361/201323177}

\bibitem[{{Sch{\"o}nrich} {et~al.}(2010){Sch{\"o}nrich}, {Binney}, \& {Dehnen}}]{schonrich2010}
{Sch{\"o}nrich}, R., {Binney}, J., \& {Dehnen}, W. 2010, \mnras, 403, 1829, \dodoi{10.1111/j.1365-2966.2010.16253.x}

\bibitem[{{Sharma} {et~al.}(2021){Sharma}, {Hayden}, \& {Bland-Hawthorn}}]{Sharma2021}
{Sharma}, S., {Hayden}, M.~R., \& {Bland-Hawthorn}, J. 2021, \mnras, 507, 5882, \dodoi{10.1093/mnras/stab2015}

\bibitem[{{Sheffield} {et~al.}(2012){Sheffield}, {Majewski}, {Johnston}, {Cunha}, {Smith}, {Cheung}, {Hampton}, {David}, {Wagner-Kaiser}, {Johnson}, {Kaplan}, {Miller}, \& {Patterson}}]{Sheffield2012}
{Sheffield}, A.~A., {Majewski}, S.~R., {Johnston}, K.~V., {et~al.} 2012, \apj, 761, 161, \dodoi{10.1088/0004-637X/761/2/161}

\bibitem[{{Skrutskie} {et~al.}(2006){Skrutskie}, {Cutri}, {Stiening}, {Weinberg}, {Schneider}, {Carpenter}, {Beichman}, {Capps}, {Chester}, {Elias}, {Huchra}, {Liebert}, {Lonsdale}, {Monet}, {Price}, {Seitzer}, {Jarrett}, {Kirkpatrick}, {Gizis}, {Howard}, {Evans}, {Fowler}, {Fullmer}, {Hurt}, {Light}, {Kopan}, {Marsh}, {McCallon}, {Tam}, {Van Dyk}, \& {Wheelock}}]{skrutskie2006}
{Skrutskie}, M.~F., {Cutri}, R.~M., {Stiening}, R., {et~al.} 2006, \aj, 131, 1163, \dodoi{10.1086/498708}

\bibitem[{{Spitoni} {et~al.}(2019){Spitoni}, {Silva Aguirre}, {Matteucci}, {Calura}, \& {Grisoni}}]{Spitoni2019}
{Spitoni}, E., {Silva Aguirre}, V., {Matteucci}, F., {Calura}, F., \& {Grisoni}, V. 2019, \aap, 623, A60, \dodoi{10.1051/0004-6361/201834188}

\bibitem[{{Spitoni} {et~al.}(2021){Spitoni}, {Verma}, {Silva Aguirre}, {Vincenzo}, {Matteucci}, {Vai{\v{c}}ekauskait{\.{e}}}, {Palla}, {Grisoni}, \& {Calura}}]{Spitoni2021}
{Spitoni}, E., {Verma}, K., {Silva Aguirre}, V., {et~al.} 2021, \aap, 647, A73, \dodoi{10.1051/0004-6361/202039864}

\bibitem[{{Spitoni} {et~al.}(2023){Spitoni}, {Recio-Blanco}, {de Laverny}, {Palicio}, {Kordopatis}, {Schultheis}, {Contursi}, {Poggio}, {Romano}, \& {Matteucci}}]{Spitoni2023}
{Spitoni}, E., {Recio-Blanco}, A., {de Laverny}, P., {et~al.} 2023, \aap, 670, A109, \dodoi{10.1051/0004-6361/202244349}

\bibitem[{{Starkenburg} {et~al.}(2017){Starkenburg}, {Martin}, {Youakim}, {Aguado}, {Allende Prieto}, {Arentsen}, {Bernard}, {Bonifacio}, {Caffau}, {Carlberg}, {C{\^o}t{\'e}}, {Fouesneau}, {Fran{\c{c}}ois}, {Franke}, {Gonz{\'a}lez Hern{\'a}ndez}, {Gwyn}, {Hill}, {Ibata}, {Jablonka}, {Longeard}, {McConnachie}, {Navarro}, {S{\'a}nchez-Janssen}, {Tolstoy}, \& {Venn}}]{Starkenburg2017}
{Starkenburg}, E., {Martin}, N., {Youakim}, K., {et~al.} 2017, \mnras, 471, 2587, \dodoi{10.1093/mnras/stx1068}

\bibitem[{Steinmetz(2002)}]{steinmetz2002}
Steinmetz, M. 2002, arXiv preprint astro-ph/0211417

\bibitem[{{Tinsley}(1980)}]{Tinsley1980}
{Tinsley}, B.~M. 1980, \fcp, 5, 287, \dodoi{10.48550/arXiv.2203.02041}

\bibitem[{{van der Kruit} \& {Freeman}(2011)}]{Kruit2011}
{van der Kruit}, P.~C., \& {Freeman}, K.~C. 2011, \araa, 49, 301, \dodoi{10.1146/annurev-astro-083109-153241}

\bibitem[{{Vasiliev}(2019)}]{Vasiliev2019}
{Vasiliev}, E. 2019, \mnras, 482, 1525, \dodoi{10.1093/mnras/sty2672}

\bibitem[{{Villalobos} \& {Helmi}(2008)}]{Villalobos2008}
{Villalobos}, {\'A}., \& {Helmi}, A. 2008, \mnras, 391, 1806, \dodoi{10.1111/j.1365-2966.2008.13979.x}

\bibitem[{{Vincenzo} {et~al.}(2019){Vincenzo}, {Spitoni}, {Calura}, {Matteucci}, {Silva Aguirre}, {Miglio}, \& {Cescutti}}]{Vincenzo2019}
{Vincenzo}, F., {Spitoni}, E., {Calura}, F., {et~al.} 2019, \mnras, 487, L47, \dodoi{10.1093/mnrasl/slz070}

\bibitem[{{Wilson} {et~al.}(2019){Wilson}, {Hearty}, {Skrutskie}, {Majewski}, {Holtzman}, {Eisenstein}, {Gunn}, {Blank}, {Henderson}, {Smee}, {Nelson}, {Nidever}, {Arns}, {Barkhouser}, {Barr}, {Beland}, {Bershady}, {Blanton}, {Brunner}, {Burton}, {Carey}, {Carr}, {Colque}, {Crane}, {Damke}, {Davidson}, {Dean}, {Di Mille}, {Don}, {Ebelke}, {Evans}, {Fitzgerald}, {Gillespie}, {Hall}, {Harding}, {Harding}, {Hammond}, {Hancock}, {Harrison}, {Hope}, {Horne}, {Karakla}, {Lam}, {Leger}, {MacDonald}, {Maseman}, {Matsunari}, {Melton}, {Mitcheltree}, {O'Brien}, {O'Connell}, {Patten}, {Richardson}, {Rieke}, {Rieke}, {Roman-Lopes}, {Schiavon}, {Sobeck}, {Stolberg}, {Stoll}, {Tembe}, {Trujillo}, {Uomoto}, {Vernieri}, {Walker}, {Weinberg}, {Young}, {Anthony-Brumfield}, {Bizyaev}, {Breslauer}, {De Lee}, {Downey}, {Halverson}, {Huehnerhoff}, {Klaene}, {Leon}, {Long}, {Mahadevan}, {Malanushenko}, {Nguyen}, {Owen}, {S{\'a}nchez-Gallego}, {Sayres}, {Shane}, {Shectman}, {Shetrone}, {Skinner}, {Stauffer}, \& {Zhao}}]{Wilson2019}
{Wilson}, J.~C., {Hearty}, F.~R., {Skrutskie}, M.~F., {et~al.} 2019, \pasp, 131, 055001, \dodoi{10.1088/1538-3873/ab0075}

\bibitem[{{Wisnioski} {et~al.}(2015){Wisnioski}, {F{\"o}rster Schreiber}, {Wuyts}, {Wuyts}, {Bandara}, {Wilman}, {Genzel}, {Bender}, {Davies}, {Fossati}, {Lang}, {Mendel}, {Beifiori}, {Brammer}, {Chan}, {Fabricius}, {Fudamoto}, {Kulkarni}, {Kurk}, {Lutz}, {Nelson}, {Momcheva}, {Rosario}, {Saglia}, {Seitz}, {Tacconi}, \& {van Dokkum}}]{Wisnioski2015}
{Wisnioski}, E., {F{\"o}rster Schreiber}, N.~M., {Wuyts}, S., {et~al.} 2015, \apj, 799, 209, \dodoi{10.1088/0004-637X/799/2/209}

\bibitem[{{Yanny} {et~al.}(2009){Yanny}, {Rockosi}, {Newberg}, {Knapp}, {Adelman-McCarthy}, {Alcorn}, {Allam}, {Allende Prieto}, {An}, {Anderson}, {Anderson}, {Bailer-Jones}, {Bastian}, {Beers}, {Bell}, {Belokurov}, {Bizyaev}, {Blythe}, {Bochanski}, {Boroski}, {Brinchmann}, {Brinkmann}, {Brewington}, {Carey}, {Cudworth}, {Evans}, {Evans}, {Gates}, {G{\"a}nsicke}, {Gillespie}, {Gilmore}, {Nebot Gomez-Moran}, {Grebel}, {Greenwell}, {Gunn}, {Jordan}, {Jordan}, {Harding}, {Harris}, {Hendry}, {Holder}, {Ivans}, {Ivezi{\v{c}}}, {Jester}, {Johnson}, {Kent}, {Kleinman}, {Kniazev}, {Krzesinski}, {Kron}, {Kuropatkin}, {Lebedeva}, {Lee}, {French Leger}, {L{\'e}pine}, {Levine}, {Lin}, {Long}, {Loomis}, {Lupton}, {Malanushenko}, {Malanushenko}, {Margon}, {Martinez-Delgado}, {McGehee}, {Monet}, {Morrison}, {Munn}, {Neilsen}, {Nitta}, {Norris}, {Oravetz}, {Owen}, {Padmanabhan}, {Pan}, {Peterson}, {Pier}, {Platson}, {Re Fiorentin}, {Richards}, {Rix}, {Schlegel}, {Schneider}, {Schreiber}, {Schwope}, {Sibley}, {Simmons},
  {Snedden}, {Allyn Smith}, {Stark}, {Stauffer}, {Steinmetz}, {Stoughton}, {SubbaRao}, {Szalay}, {Szkody}, {Thakar}, {Sivarani}, {Tucker}, {Uomoto}, {Vanden Berk}, {Vidrih}, {Wadadekar}, {Watters}, {Wilhelm}, {Wyse}, {Yarger}, \& {Zucker}}]{SEGUE1}
{Yanny}, B., {Rockosi}, C., {Newberg}, H.~J., {et~al.} 2009, \aj, 137, 4377, \dodoi{10.1088/0004-6256/137/5/4377}

\bibitem[{{York} {et~al.}(2000){York}, {Adelman}, {Anderson}, {Anderson}, {Annis}, {Bahcall}, {Bakken}, {Barkhouser}, {Bastian}, {Berman}, {Boroski}, {Bracker}, {Briegel}, {Briggs}, {Brinkmann}, {Brunner}, {Burles}, \& {SDSS Collaboration}}]{york2000}
{York}, D.~G., {Adelman}, J., {Anderson}, Jr., J.~E., {et~al.} 2000, \aj, 120, 1579.
\newblock \url{http://adsabs.harvard.edu/abs/2000AJ....120.1579Y}

\bibitem[{{Yoshii}(1982)}]{Yoshii1982}
{Yoshii}, Y. 1982, \pasj, 34, 365

\bibitem[{{Yoshii} {et~al.}(1987){Yoshii}, {Ishida}, \& {Stobie}}]{Yoshii1987}
{Yoshii}, Y., {Ishida}, K., \& {Stobie}, R.~S. 1987, \aj, 93, 323, \dodoi{10.1086/114317}

\bibitem[{{Zhao} {et~al.}(2012){Zhao}, {Zhao}, {Chu}, {Jing}, \& {Deng}}]{LAMOST2012}
{Zhao}, G., {Zhao}, Y.-H., {Chu}, Y.-Q., {Jing}, Y.-P., \& {Deng}, L.-C. 2012, Research in Astronomy and Astrophysics, 12, 723, \dodoi{10.1088/1674-4527/12/7/002}

\end{thebibliography}

%\appendix

\appendix
%\label{sec:discussion} 

\section{Age prior of the disk}
\label{sec:appendix} 

\begin{figure*}[ht!]
    \centering
    \includegraphics[width=\columnwidth]{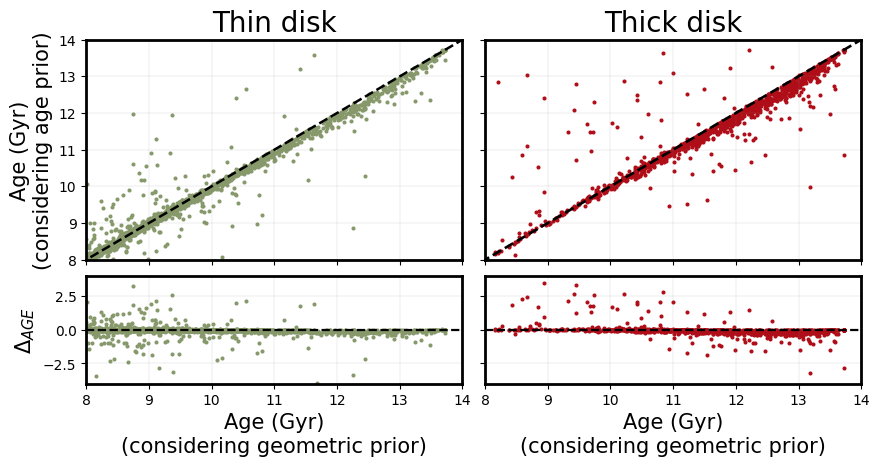}
    \caption{Estimated stellar ages for thin (leftmost) and thick disk (rightmost panel) stars older than 8\,Gyr considering the age prior on the vertical axis and considering only the spatial distribution of stars with no age prior on the horizontal axis. The lower panels show the difference between the two calculated ages ($\Delta_{\text{AGE}}$).}
    \label{prior}
\end{figure*}

The \texttt{StarHorse} ages used in this paper, as presented in \citet{Queiroz2023}, assume Gaussian age priors, following the methodology described in \citet{Queiroz2018, Queiroz2020, Queiroz2023}. Consequently, the ages of stars in the thin and thick disk populations analyzed here are influenced by this prior. To verify that this does not bias our main results, we repeated the analysis using \texttt{StarHorse} ages derived for the same APOGEE DR17 sample but without the age prior, relying only on a geometric prior. Figure \ref{prior} compares the ages obtained with (vertical axis) and without (horizontal axis) the age prior for the thin and thick disk populations, focusing on ages greater than 8 Gyr. Notably, the age prior tends to make older stars appear younger compared to estimates based solely on spatial position. This test confirms that the presence of an ancient thin disk population is not an artifact of the prior. Since we identify low-$\alpha$ stars with dynamically cooler orbits and ages exceeding 11 Gyr (with uncertainties under 1 Gyr), even though the age prior systematically shifts ages downward, our results remain robust against this effect.

\section{The Thin and Thick disk's rotational support}
\label{sec:appendix_B} 

\begin{figure}
    \centering
    \includegraphics[width=\columnwidth]{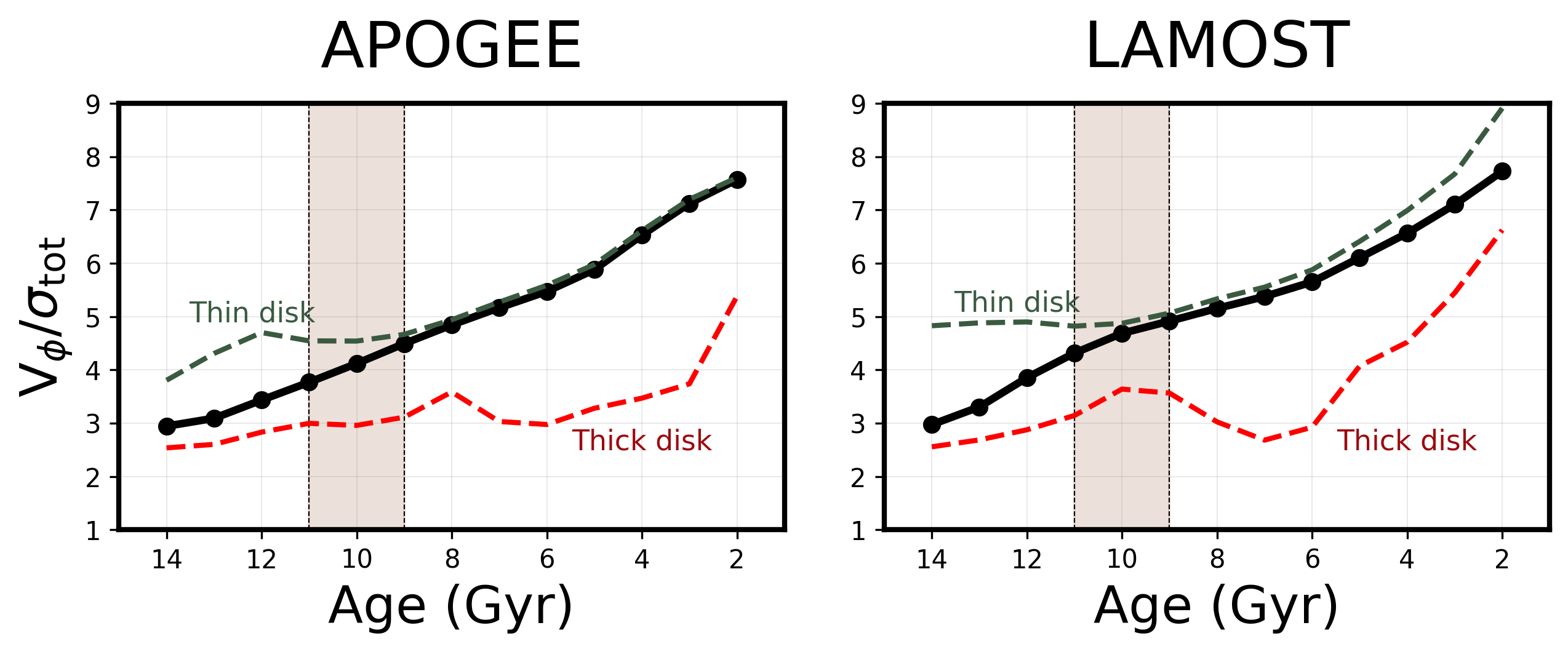}
    \caption{$V_{\phi}/\sigma_{\text{tot}}$ as a function of age is represented by the black line, where $\sigma_{\text{tot}} = \sqrt{\sigma_{V_r}^2 + \sigma_{V_z}^2 + \sigma_{V_{\phi}}^2}$. We consider age bins of 1\,Gyr. APOGEE and LAMOST samples are represented in the upper and lower panels, respectively. The dashed lines represent the individual samples from the thin (green) and thick (red) disks.}
    \label{vphi_sigma}
\end{figure}

A good measure of disk rotation is the amount of stellar rotation versus velocity dispersion, $V_{\text{rot}}/\sigma_{\text{tot}}$ (e.g., \citealt{Kruit2011}), where the threshold $V_{\text{rot}}/\sigma_{\text{tot}} = 1$ has been used to distinguish ‘dispersion dominated’ ($V_{\text{rot}}/\sigma_{\text{tot}} < 1$, as bulges, elliptical and dwarf spheroidal) and ‘rotation dominated’ ($V_{\text{rot}}/\sigma_{\text{tot}} > 1$, disk) galaxies (\citealt{Wisnioski2015, Johnson2018}). Figure \ref{vphi_sigma} shows the evolution of $V_{\phi}/\sigma_{\text{tot}}$ as a function of age for APOGEE and LAMOST samples in the top and bottom panels, respectively, and the dashed lines represent the individual samples from the thin (green) and thick (red) disks. Here, we use $V_{\text{rot}} = V_{\phi}$ and $\sigma_{\text{tot}} = \sqrt{\sigma_{V_r}^2 + \sigma_{V_z}^2 + \sigma_{V_{\phi}}^2}$. Our samples are contained within $V_{\text{rot}}/\sigma_{\text{tot}} > 1$, as expected for the disk structure, and follow an increasing trend with age, including for individual thin and thick disc samples. As shown previously in Figure \ref{std} and reinforced by Figure \ref{vphi_sigma}, the old population of the thin disk has lower rotational support than the young part, but it still reaches higher values than the thick disk for ages $>$ 11\,Gyr.

%\clearpage

%\setcounter{table}{0}

\end{document}